\documentclass[iop]{emulateapj}
\usepackage{graphics}
\usepackage{epsfig}
\usepackage{booktabs}
\usepackage{rotating}
\usepackage{fancyhdr}
\usepackage{threeparttable}
\usepackage{amsmath}
\usepackage{color}

\newcommand{\ltsimeq}{\la}
\newcommand{\gtsimeq}{\ga}

\newcommand{\msun}{M$_{\odot}$}

\newcommand{\HII}{H~{\sc ii}}

\accepted{May 2015}
\shortauthors{McQuinn et~al.}
\shorttitle{Calibrating UV Star Formation Rates for Dwarf Galaxies from STARBIRDS}

\begin{document}
\title{Calibrating UV Star Formation Rates for Dwarf Galaxies from STARBIRDS
\footnote{Based on observations made with the NASA/ESA Hubble Space Telescope, and obtained from the Hubble Legacy Archive, which is a collaboration between the Space Telescope Science Institute (STScI/NASA), the Space Telescope European Coordinating Facility (ST-ECF/ESA) and the Canadian Astronomy Data Centre (CADC/NRC/CSA).}}
\author{Kristen~B.~W. McQuinn\altaffilmark{1}, 
Evan D.~Skillman\altaffilmark{1},
Andrew E.~Dolphin\altaffilmark{2},
Noah P.~Mitchell\altaffilmark{1,3,4}
}

\altaffiltext{1}{Minnesota Institute for Astrophysics, University of Minnesota, 116 Church Street, S.E., Minneapolis, MN 55455, \ {\it kmcquinn@astro.umn.edu}} 
\altaffiltext{2}{Raytheon Company, 1151 E. Hermans Road, Tucson, AZ 85756}
\altaffiltext{3}{Department of Physics, St.~Olaf College, 1520 St. Olaf Ave, Northfield, MN 55057}
\altaffiltext{4}{Current address: Department of Physics and the James Frank Institute, University of Chicago, 929 East 57th Street, Chicago, Illinois 60637}

\begin{abstract}
Integrating our knowledge of star formation traced by observations at different wavelengths is essential for correctly interpreting and comparing star formation activity in a variety of systems and environments. This study compares extinction corrected integrated ultraviolet (UV) emission from resolved galaxies with color-magnitude diagram (CMD) based star formation rates (SFRs) derived from resolved stellar populations and CMD fitting techniques in 19 nearby starburst and post-starburst dwarf galaxies. The datasets are from the panchromatic STARBurst IRregular Dwarf Survey (STARBIRDS) and include deep legacy GALEX UV imaging, HST optical imaging, and Spitzer MIPS imaging. For the majority of the sample, the integrated near UV fluxes predicted from the CMD-based SFRs - using four different models - agree with the measured, extinction corrected, integrated near UV fluxes from GALEX images, but the far UV predicted fluxes do not. Further, we find a systematic deviation between the SFRs based on integrated far UV luminosities and existing scaling relations, and the SFRs based on the resolved stellar populations. This offset is \textit{not} driven by different star formation timescales, variations in SFRs, UV attenuation, nor stochastic effects. This first comparison between CMD-based SFRs and an integrated FUV emission SFR indicator suggests that the most likely cause of the discrepancy is the theoretical FUV-SFR calibration from stellar evolutionary libraries and/or stellar atmospheric models. We present an empirical calibration of the FUV-based SFR relation for dwarf galaxies, with uncertainties, which is $\sim$53\% larger than previous relations.
\end{abstract}


\section{Star Formation Rate Measurements\label{intro}}
Measurements of the star formation histories (SFHs) of galaxies are a key to understanding the evolution of galaxies as a function of redshift. Qualitatively, the SFH can be deduced by classifying a galaxy as hosting an older, red population, a younger blue population, or both. Quantitatively, the SFH often reduces to a single measurable star formation rate (SFR) occurring in a galaxy at a particular redshift. Significant effort has been put into calculating such SFRs using a variety of techniques. One of the simplest methods is based on measuring the integrated light of a galaxy in a given wavelength regime (i.e., H$\alpha$, UV, or infrared (IR) wavelengths), and converting this flux to a SFR using theoretically derived scaling relations \citep[e.g.,][]{Madau1998, Kennicutt1998, Calzetti2007, Calzetti2010, Hao2011, Murphy2011}. These SFR scaling relations are based on assumptions about the origin of the integrated light. For example, UV emission is assumed to originate from stellar populations with a well-defined range in age (and stellar mass), H$\alpha$ emission is assumed to originate from gas ionized by massive stars younger than $\sim10$ Myr, and light in the mid-IR regime is assumed to be emitted by dust that has been primarily heated by relatively young stars. 

Calculating SFRs using simple scaling relations is most robust in the UV regime. In contrast to the H$\alpha$ and IR emission that originates from gas and dust heated by star formation, the UV emission originates from the photospheres of young stars of intermediate and high mass (M$\gtsimeq3$ \msun) and, thus, is a direct tracer of the recent star formation. SFRs derived from integrated emission measurements are dependent on the star formation timescales probed by a particular wavelength regime. The stars responsible for producing the majority of the UV flux have a lifespan of $\sim$100 Myr, versus $\sim5-10$ Myr for the more massive stars (M$\gtsimeq17$ \msun) responsible for ionizing hydrogen and producing the secondary H$\alpha$ emission. Calculation of SFRs based on integrated UV fluxes have been shown to be more accurate at lower values than SFRs based on integrated H$\alpha$ fluxes, which can be suppressed due to the incomplete sampling of the high mass end of the initial mass function (IMF) at low SFRs \citep[e.g.,][]{Boselli2009, Goddard2010, Koda2012}. Similarly, temporal stochasticity can impact the interpretation of integrated UV emission measurements, although to a lesser extent \citep[e.g.,][]{Lee2009, Lee2011, Weisz2012} as the far ultraviolet (FUV) emission is produced by a wider range of stellar masses than the H$\alpha$ emission. 

However, interpretation of the FUV emission can be complicated. Recent models of star formation (SF) in low-mass galaxies have found that different factors such as the total mass of a galaxy, the degree of clustering, and the overall level of activity may impact the total FUV emission produced at a given SFR \citep{Fumagalli2011, daSilva2012, daSilva2014}. Further, the theoretically defined scaling relations are based on the assumption that the UV emission originates in a system with a constant SFR. While this may be a valid approximation for spiral galaxies with a large number of independent star forming regions, this assumption can break down on smaller spatial scales or in low-mass systems. Similar to H$\alpha$ emission, UV emission can be significantly affected by extinction, which can result in large uncertainties in measured SFRs. Best practices to accurately account for this attenuation require measurements of the energy absorbed by the dust and re-emitted at infrared wavelengths. Empirical studies using IR emission to trace dust content and approximate energy balancing schemes to convert this emission to an equivalent UV luminosity have improved corrections for UV attenuation \citep[e.g.,][]{Buat2005, Calzetti2007, Kennicutt2009}. 

The Galaxy Evolutionary Explorer (GALEX) mission \citep{Martin2005} surveyed two-thirds of the sky in the UV, greatly benefiting extragalactic studies in the UV. GALEX observations have been adept at characterizing star formation in low density gas environments, such as the outer disks of spiral and dwarf Irregular galaxies \citep{Thilker2007}, providing a new means for tracing recent star formation in more quiescent environments. The ease of measuring the total integrated UV emission from GALEX images for a significant number of galaxies has made accurate calibration of UV-based SFR scaling relations even more important.

Theoretical SFR relations based on scaling integrated UV emission measurements are derived and calibrated from synthetic spectral models of single-aged stellar populations \citep[e.g.,][]{Bruzual1993, Leitherer1999}. They are based on the premise that if a SFR has been constant over a $\sim100$ Myr timescale, the birth of stars that significantly contribute to the FUV emission is in equilibrium with the death of similar UV contributors. Thus, the UV emission is proportional to the SFR itself. UV scaling relations are widely used because of their relative simplicity and because they have been consistently calibrated against each other \citep[e.g.,][and references therein]{Kennicutt2012}. The most widely used UV scaling relation, presented in \citet{Kennicutt1998}, is based on the theoretical results from \citet{Madau1998} over the wavelength range 1500-2800 \AA. Adjustments to the scaling relation have been made using the stellar evolution models of the Starburst99 code \citep{Leitherer1999}. In one study, the scaling factor was raised by 14\% for the same IMF and metallicity \citep{Iglesias-Paramo2006}, but it is unclear what caused the increase. In a separate study, the scaling factor was lowered by 10\% based on the lower metallicity values of the galaxy sample \citep{Hunter2010}. More recently, the theoretical relation was re-calibrated using updated stellar evolution libraries and spectral models \citep{Hao2011, Murphy2011} with the Starburst99 code \citep{Leitherer1999, Vazquez2005}. The resulting revision is $\sim5$\% lower than the original value from \citet{Kennicutt1998} for the same IMF. 

The FUV-based SFR scaling relation was empirically calibrated to be 30\% lower than the theoretical relation from \citet{Kennicutt1998} for the GALEX FUV bandpass using spectral energy distribution fitting of integrated light measurements for a large sample of high-mass galaxies with a mean metallicity of 0.8 Z$_{\odot}$ \citep{Salim2007}. This study reported that calibration of this GALEX FUV-based SFR scaling relation was difficult for lower mass systems as proper treatment of the UV attenuation using only FUV and near ultraviolet (NUV) magnitudes was unreliable at lower SFRs.  Further, the empirical calibration was derived from a sample of galaxies with a variety of SFHs. Thus, the assumption of constant star formation over the lifetime of UV emitting stars is compromised. 

Despite both the theoretical and empirical adjustments, what is still lacking is an independent, empirical calibration of the UV scaling relation in a sample with well-measured and well-constrained star formation properties outside of the synthesis model approach. A different approach to measuring SFRs in nearby galaxies is to derive star formation histories (SFHs, or SFRs(t,Z)) from resolved stellar populations \citep[e.g.,][]{Dolphin1997, Tosi1989, Tolstoy1996, Gallart1996, Holtzman1999, Harris2001}. This method uses CMD fitting techniques to match observed CMDs to composite synthetic stellar populations generated from stellar evolution libraries. The best-fit modeled CMD represents the most likely SFH of the system. While deriving SFRs at the earliest epochs using this technique requires deep optical imaging reaching below the oldest main sequence turn-off, recent SFRs can be robustly derived from well-populated CMDs of less photometric depth which include significant numbers of upper main sequence and blue helium burning stars \citep{Dolphin2002, McQuinn2010b}. 

SFHs derived from resolved stellar populations provide an independent measure of the SF activity which can be compared with SFRs from integrated UV measurements. In low-mass systems, which may be vulnerable to stochastic effects, the resolved stars also provide constraints on such effects. Such a comparison requires a sample of nearby galaxies with both significant levels of recent star formation and low levels of extinction. Thus, nearby starburst dwarf galaxies make an ideal sample. By definition, starbursts have elevated levels of recent star formation activity, making them UV bright systems. Previously thought to have durations of only $\sim10$ Myr \citep[e.g.,][]{Schaerer1999, Mas-Hesse1999, Thornley2000, Tremonti2001}, \citet{McQuinn2009,McQuinn2010b} have argued that starbursts in dwarfs are a longer-lived phenomenon, lasting $>100$ Myr, based on a comparison of their recent star-formation activity with their historical averages derived from resolved stellar populations CMD-fitting techniques. From a companion study of the same data, the spatial extent of the star-formation activity was observed to range from centrally concentrated to widely distributed \citep{McQuinn2012}. The longer lived, often spatially distributed starbursts approximate a constant SFR over UV timescales - an assumption that underlies the SFR scaling relations. Dwarf galaxies also have lower metallicities than more massive systems, with correspondingly lower UV attenuation from dust. This reduces the uncertainties and complexities in applying extinction corrections to the UV emission measurements. 

In this study, we have undertaken a comprehensive comparison of the integrated UV emission from star formation based on deep GALEX imaging with the resolved stellar populations imaged with the Hubble Space Telescope (HST) in a sample of nearby starburst and post-starburst dwarf galaxies. These observations are part of the panchromatic STARBurst IRregular Dwarf Survey (STARBIRDS) presented in a companion paper \citep{McQuinn2015}, and have been used in previous studies to measure the characteristics of the starburst phenomenon \citep{McQuinn2009, McQuinn2010a, McQuinn2010b, McQuinn2012}. Although internal extinction was expected to be low at the metallicities of the sample, we applied the current best practice to correct for UV extinction using the infrared flux measured from Spitzer Space Telescope MIPS imaging. The data are discussed in Section~2. In Sections~3 and 4 of this study, we test our understanding of the connection between optical and UV emission from stellar populations. Using four different models and the SFHs reconstructed from the resolved stellar populations, we predict the NUV and FUV fluxes from the galaxies, and compare these values with the extinction corrected, observed values of UV emission. This comparison provides a critical test for our understanding of the optical and UV emission in galaxies because it uses an independent measurement of the star formation from the resolved stellar populations to model the UV flux and then compares this modeled flux with the measured emission from observations. Moreover, it provides a check on a number of the parameters including whether or not the SFHs are accurately describing the star formation activity in the galaxies, whether stochastic effects are dominating the results in low-mass galaxies, and whether the UV emission has been accurately corrected for extinction. In Section~5, we compare the SFHs over timescales applicable to UV bright stars from resolved stellar populations with SFRs based on integrated FUV fluxes. In Section~6 we present an empirical calibration of the integrated FUV emission SFR scaling relation for dwarf galaxies based on the resolved stellar populations and in Section~7 we discuss possible sources of the dispersion in SFRs based on integrated FUV emission. Our conclusions are summarized in Section~8. In the Appendix, we provide a detailed discussion of the possible role observational uncertainties have on a comparison between predicted and measured UV fluxes.

\section{The Galaxy Sample and Observations\label{obs}}
Table~\ref{tab:galaxies} lists the 19 nearby (D$<6$ Mpc) starburst and post-starburst dwarf galaxies in the STARBIRDS sample and their various physical properties. The sample was adopted from previous studies deriving the SFHs of nearby starburst dwarf galaxies using HST F606W (or F555W) and F814W archival imaging \citep{McQuinn2010a}. The sample was defined based on data from the HST archive that met the following criteria: First, the galaxies had to be close enough that the HST imaging instruments could resolve the stellar populations. Second, both V-band (i.e., F555W or F606W) and I-band (i.e., F814W) observations were available for each system. Third, the I-band photometric depth reached a minimum of $\sim2$ mag below the tip of the red giant branch. The galaxies span a wide parameter space in SFR, metallicity, and physical size, including both stronger, well-known starburst dwarf galaxies in the nearby universe (e.g., NGC~1569, NGC~4449, NGC~5253), smaller, less well-known starbursts (e.g., NGC~4068, UGC~6456), and post-starburst galaxies (Antlia, UGC~9128, NGC~4163, NGC~6789, NGC~625). Note, however, that it is not a volume limited sample. A more detailed discussion of the sample selection and its limitations can be found in \citet[][see Section 2.1]{McQuinn2015}.

The data consist of observations from the GALEX, HST, and Spitzer space telescopes. All reduced images are publically available through the STARBIRDS data archive available as high-level science products through the Mikulski Archive for Space Telescopes (MAST). The data reduction procedures are described in detail in a companion paper \citep{McQuinn2015}. Additional information on the processing of the HST observations can be found in \citet{McQuinn2010a}. Below, we briefly summarize the data reduction process for each dataset.

\subsection{GALEX UV Observations}
\begin{figure}
\plottwo{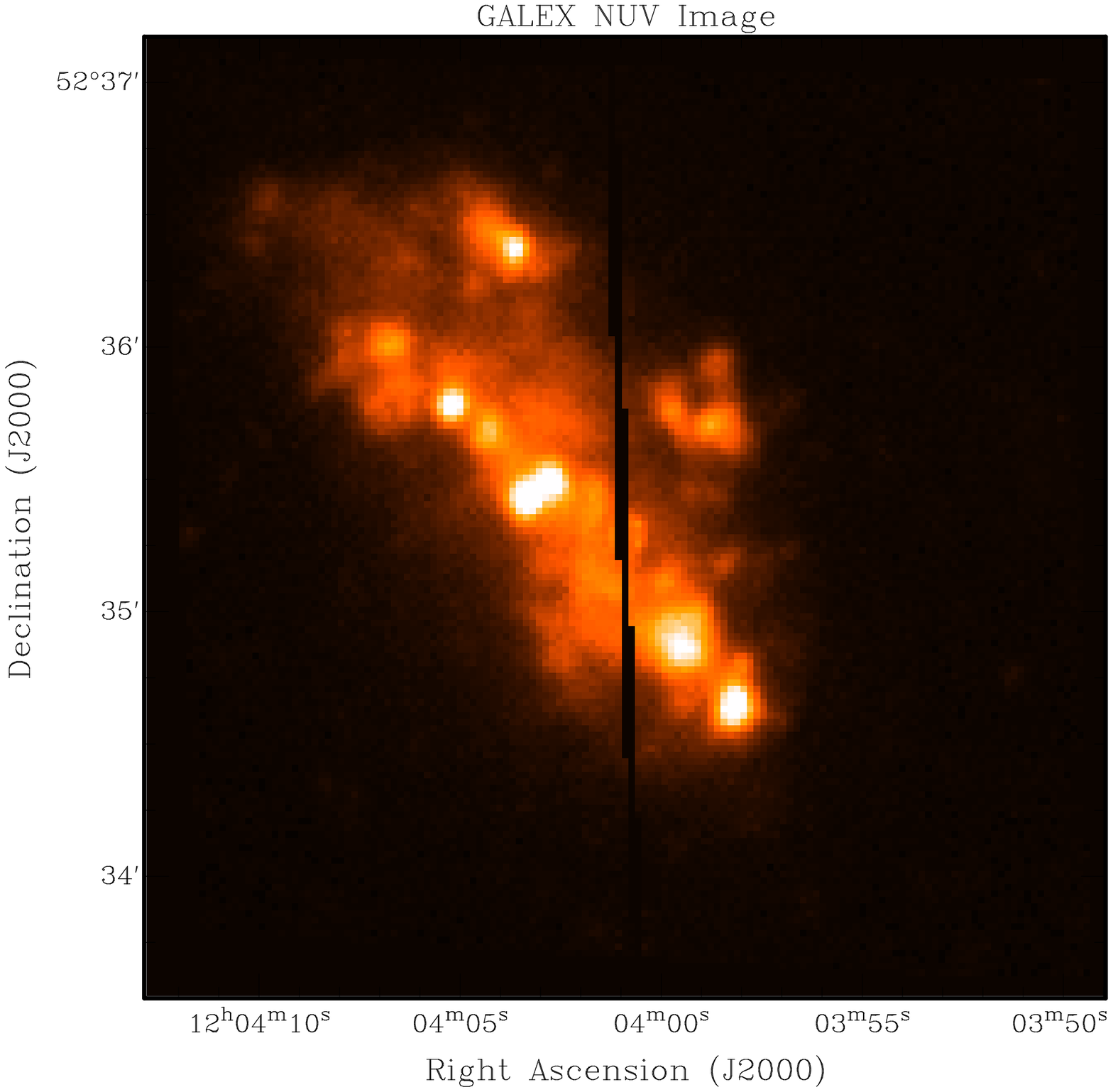}{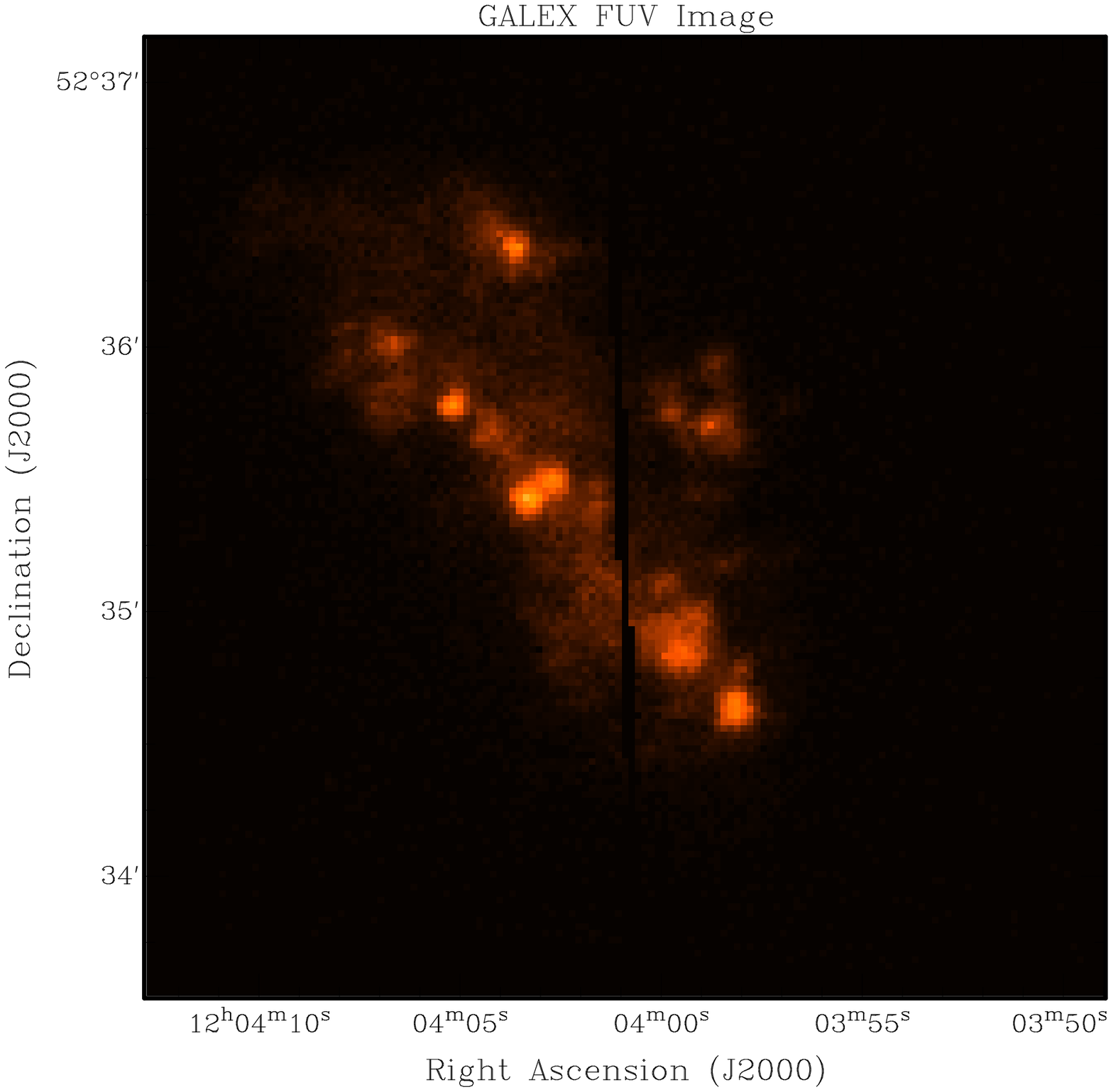}
\caption{Representative of our sample, GALEX NUV and FUV images of NGC~4068 are shown in the left and right panels respectively with North up and East left. The field of view matches the footprint of the HST ACS images, and excludes the chip gap, which facilitates a direct comparison of the star formation imaged by the different instruments. Typical of the sample, the image of NGC~4068 shows several knots of UV emission tracing the star forming regions, as well as fairly ubiquitous diffuse UV emission. The NUV image is significantly deeper than the FUV image and traces more of the lower surface brightness, diffuse UV emission.}
\label{fig:galex_image}
\end{figure}

The GALEX instrument, with both FUV ($\lambda\ 1350 - 1750$ \AA) and NUV ($\lambda\ 1750 - 2750$ \AA) bandpasses, is described in detail in \citet{Martin2005}, and its in-orbit performance and calibration by \citet{Morrissey2005, Morrissey2007}. The GALEX data used here include NUV and FUV images from multiple observing programs. The data were processed through the GALEX pipeline (v7.1) and all observations were co-added into one FUV and one NUV image per target. Figure~\ref{fig:galex_image} presents an example of the GALEX NUV and FUV images for the starburst galaxy NGC~4068, cropped to the field of view of HST observations of this system. In both images, UV bright knots from star forming regions are apparent, as well as extended, diffuse emission. The NUV image is considerably deeper than the FUV image and, thus, is sensitive to lower surface brightness features in the diffuse emission.

The observations used to create the deeper, co-added tiles can have slightly varying fields of view. Therefore, the GALEX pipeline provides high resolution relative response (rrhr) maps which measure the depth of the observations per pixel. In addition, the GALEX pipeline converts the GALEX satellite telemetry data (i.e., photon lists) into calibrated intensity maps in counts per second and performs background subtraction on the co-added images. We used these background subtracted intensity maps for the lower surface brightness galaxies, but found that these images over-subtracted the background for higher surface brightness regions in 5 galaxies (NGC~625, Ho~II, NGC~6822, NGC~4214, NGC~4449) and for one lower surface brightness galaxy affected by cirrus contamination (Antlia). In these cases, we performed a custom background subtraction. 

As we are interested in directly comparing the UV emission with the optical HST images, we cropped the larger, circular $1.25^{\circ}$ GALEX field of view to the smaller footprint of the HST images. Finally, we custom masked the foreground and background contamination  in the cropped, background subtracted intensity maps using the IRAF$\footnote{IRAF is distributed by the National Optical Astronomy Observatory, which is operated by the Association of Universities for Research in Astronomy (AURA) under cooperative agreement with the National Science Foundation.}$ task $\textsc{APPHOT}$ task {\tt imedit}. The integrated NUV and FUV fluxes were measured in these post-masked images using the IRAF aperture photometry functionality. The fluxes are reported in \citet{McQuinn2015}, and repeated here in Table~\ref{tab:flux_lum} for convenience. 

\subsection{HST Optical Observations}
\begin{figure}
\includegraphics[width=0.45\textwidth]{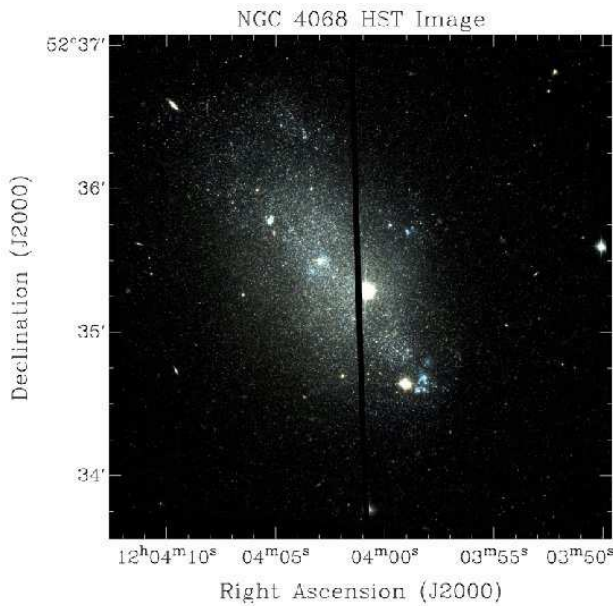}
\caption{Representative of our sample, this HST composite image of NGC~4068 was created using the F606W image (blue), F814W image (red), and an averaged F606W and F814W image (green) with North up and East left. As the HST optical imaging provides a direct census of the resolved stellar populations, the HST footprint defined the field of view for comparing the star formation imaged by the different instruments. Apparent in the image are several blue star forming regions which correspond to the knots of UV emission seen in Figure~\ref{fig:galex_image}. An underlying redder, older stellar population is seen extending to larger radii. }
\label{fig:hst_image}
\end{figure}

The HST observations are comprised of Advanced Camera for Surveys \citep[ACS;][]{Ford1998} and Wide Field Planetary Camera 2 \citep[WFPC2;][]{Holtzman1995} archival images obtained with the F814W filter and either the F555W or F606W filter. The HST data were processed and cleaned by the standard HST pipeline. Figure~\ref{fig:hst_image} presents an example of the HST images for NGC~4068, the same galaxy shown in Figure~\ref{fig:galex_image}. This HST composite image shows the resolved stellar populations including a young, blue stellar population and an underlying older, red stellar population. 

Photometry was performed on the ACS images using DOLPHOT \citep{Dolphin2000} and on the WFPC2 images using HSTphot \citep{Dolphin2000}. Artificial star tests were performed to measure completeness due to the finite photometric depth and crowding as well as the noise characteristics of the data using the same software. The resulting photometry star lists were uniformly processed, ensuring reliable comparison between galaxies. More details on the photometry can found in \citet{McQuinn2010a}.

\begin{figure}
\includegraphics[width=0.45\textwidth]{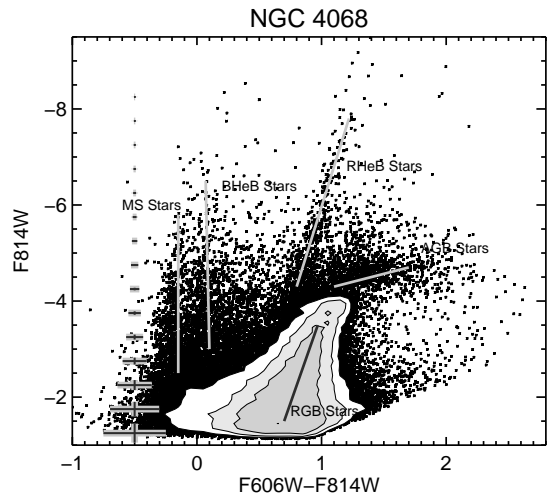}
\caption{Representative of our sample, this CMD is for NGC~4068 with average photometric uncertainties per magnitude bin shown at the left. The contours show the density of points where they would otherwise saturate the plot. The main branches of stellar evolution are noted. One of the distinguishing features in the CMDs of starburst galaxies is the notable population of red and blue helium burning stars and upper main sequence stars; both are evidence of significant recent star formation activity.}
\label{fig:hst_cmd}
\end{figure}

In Figure~\ref{fig:hst_cmd}, we show a CMD created from the photometry of the HST images for NGC~4068, which provides a direct census of the stellar populations. Evident in this representative CMD are distinct evolutionary branches of the stellar populations including the main sequence (MS), blue and red helium burning stars (BHeB and RHeB), the red giant branch (RGB), and asymptotic branch stars (AGB). Average uncertainties per magnitude bin, including uncertainties measured from completeness tests, are also shown.

\subsection{Spitzer IR Observations}
The Spitzer observations of the entire sample are comprised of archival mid-IR observations from the MIPS instrument \citep{Rieke2004,Engelbracht2007} in the 24, 70, and 160 $\micron$ bandpasses. Emission at these mid-IR wavelengths traces the dust content heated by the stellar population and can be extrapolated to estimate the dust attenuation of the UV emission. Most of the observations were obtained as part of either the Spitzer Infrared Nearby Galaxy Survey (SINGS) \citep{Kennicutt2003, Regan2004} or the Local Volume Legacy (LVL) \citep{Dale2009} projects. These data were cleaned, background subtracted, and made publically available as value-added data products of the SINGS and LVL projects. Detailed descriptions of the data processing steps are provided by these legacy programs. Four galaxies (IC~4662, NGC~1569, NGC~6789, UGC~456) were not observed in either of these surveys. For these systems, we used the MIPS observations available from the Spitzer Heritage Archive (SHA). These archival observations covered only 2 of the 3 MIPS bandpasses, and thus provide only lower limits of the dust attenuation of UV emission. We cleaned and mosaicked the SHA observations using the Spitzer-provided MOPEX software \citep{Makovoz2005a,Makovoz2005b}. Background subtraction was performed using an average flux value from relatively blank regions of the observations. 

The MIPS images were treated in the same way as the GALEX images. First, the MIPS observations were cropped to match the field of view of the HST images. Second, we custom masked the foreground and background contamination in the same manner as the GALEX UV images. Finally, the integrated mid-IR fluxes were measured in the post-masked images using the IRAF task $\textsc{APPHOT}$. The fluxes are reported in \citet{McQuinn2015}, and repeated here in Table~\ref{tab:flux_lum} for convenience. 

\begin{figure}
\includegraphics[width=0.5\textwidth]{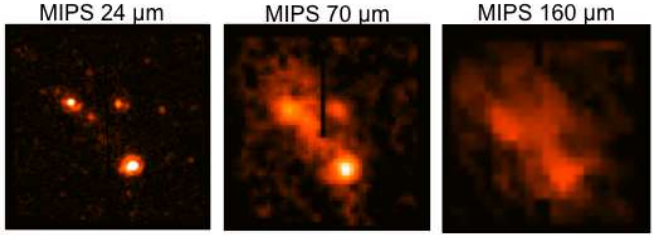}
\caption{Representative of our sample, MIPS 24~$\micron$, 70~$\micron$, and 160~$\micron$ images of NGC~4068 are shown from left to right with North up and Each left. The MIPS images are used to estimate the total IR flux from the galaxy and the corresponding UV attenuation. The field of view matches the footprint of the HST ACS image, excluding the chip gap, enabling a direct areal match of the dust emission with the UV emission measurements. The MIPS images show a few knots of 24$\micron$ emission and more extended emission at the 70$\micron$ and 160$\micron$ wavelengths. The generally low level of IR emission is typical of this low metallicity sample.}
\label{fig:mips_image}
\end{figure}

Figure~\ref{fig:mips_image} presents an example of the MIPS images at 24~$\micron$, 70~$\micron$, and 160~$\micron$ for NGC~4068, the same galaxy shown in Figures~\ref{fig:galex_image}$-$\ref{fig:hst_cmd}. Typical of most of the sample, the MIPS images show only a low level of emission, with little emission detected at 160~$\micron$. Since the majority of the sample has relatively low values of metallicity, the mid-infrared emission from dust heated by star formation is also relatively low. The MIPS data are used to estimate the total IR flux and corresponding UV attenuation. 

\section{Deriving SFHs from Resolved Stars, Predicting UV Fluxes, and Calculating SFRs from Integrated FUV Fluxes Using Existing Scaling Relations\label{lum_sfrs}}
Our calculations follow three stages. First, we describe the derivation of the SFHs from the resolved stellar populations. Second, we use four models in conjunction with the derived SFHs, to predict the integrated NUV and FUV fluxes in our sample. Third, we convert the UV and MIPS fluxes to extinction corrected UV luminosities and apply existing scaling relations to convert the FUV luminosities to SFRs. A discussion of these results, including comparisons between the measured and predicted quantities, is presented in Sections~4 and 5.

\subsection{Reconstructing SFHs from Resolved Stellar Populations}
The SFHs were previously reconstructed by \citet{McQuinn2009,McQuinn2010a}. Briefly, the SFR(t,Z) was derived by applying a CMD fitting technique \citep{Dolphin2002} with stellar evolutionary models to the photometry and artificial star lists. We used the Padova isochrones \citep{Marigo2008} because they cover the range in metallicity of our sample and include isochrones for the high mass stars that are of interest to us. The CMD fitting technique fits for distance, extinction, star formation history, and chemical enrichment history, with the metallicity constrained to increase with time (except for Antlia and NGC~2366 where the photometry reached a full magnitude below the red clump). A Salpeter IMF \citep{Salpeter1955} with mass limits of 0.1$-$120 \msun\ is assumed in the derivation. A single power-law IMF is sufficient for our purposes as the stars of interest are all high-mass, UV producing stars. The distances fit by the method were in good agreement with tip of the red giant branch (TRGB) distances from the literature, which provides a consistency check on the CMD fit. 

\begin{figure}
\includegraphics[width=0.48\textwidth]{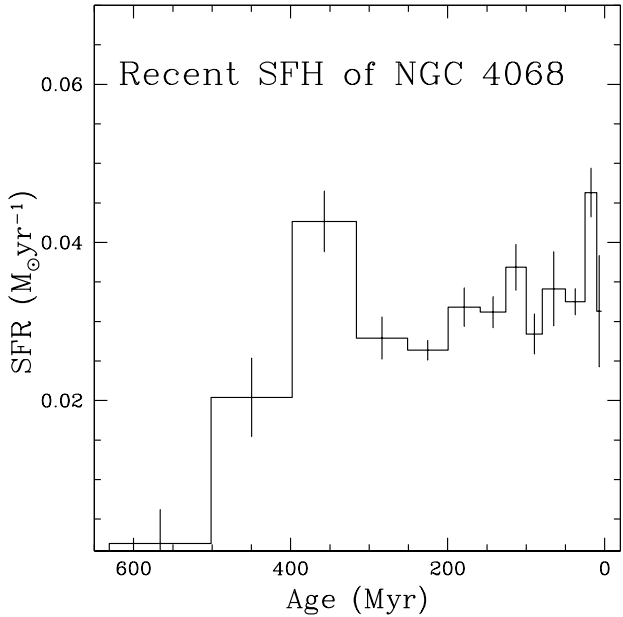}
\caption{The SFH of NGC~4068 over the last 600 Myr reconstructed from the resolved stellar populations using the CMD fitting method {\tt MATCH} \citep{Dolphin2002} and stellar evolutionary models \citep{Marigo2008}. The time resolution improves at recent times, with time bins as short as 20 Myr. The SFRs are elevated over the past $\sim500$ Myr, reflecting the starburst activity in this system. While the SFH of each galaxy is unique, SFRs elevated for more than 100 Myr is a common feature in the sample.}
\label{fig:sfh}
\end{figure}

In Figure~\ref{fig:sfh}, we reproduce the SFR(t) over the $\sim0.5$ Gyr for NGC~4068 from \citet{McQuinn2010b}. Note the elevated SFR at recent times indicative of the starburst and representative of the SFHs derived for the sample. The uncertainties in the SFR(t) include both systematic and statistical uncertainties derived from Monte Carlo simulations. The time binning reflects the inherent time resolution of the data. At older look-back times, the SFRs are not well constrained without photometry that reaches the oldest MS turn-off. However, at the more recent times shown here, the low photometric errors of the massive, young MS and HeB stars enable robust measurements of the SFRs with a much higher temporal resolution. We refer the reader to \citet{McQuinn2010a} for a full discussion of the SFHs, the time binning, and the calculation of uncertainties. 

The SFRs(t) from \citet{McQuinn2010a} were used to calculate the average SFRs over the past 100 Myr. These CMD-based SFRs are listed in Table~\ref{tab:lum_sfrs}. We chose this timescale because 90\% of the FUV emission in galaxies with a constant SFR originates from stars less than 100 Myr old \citep{Hao2011}. This temporal baseline changes to 200 Myr when considering NUV emission \citep{Hao2011}. To conservatively estimate uncertainties, we report the standard deviation in the SFRs over the past 50, 100, and 150 Myr. In all cases, the standard deviations are larger than the measured uncertainties from the SFH. The benefit of taking this approach is that it provides a measure of how constant the SFRs are in each galaxy over the timescale of interest. 

\subsubsection{Special Cases}
As with any SFR measurement, a CMD-based SFR depends on the observations to accurately trace the star-formation activity. The high-resolving power of the HST and the application of artificial star tests means that CMD-fitting techniques can successfully measure the recent star-formation in galaxies out to several Mpc. However, we have identified four galaxies in our sample which host young, massive clusters that are unresolved with the HST imaging. Thus, the CMD-based SFRs may be under-reporting the true star-formation activity. Further, due to the significant number of very young stars (t$<10$ Myr), the assumption of constant star-formation over the UV timescale is compromised. Thus, the FUV-based SFRs may be over-reporting the true star-formation activity.

Specifically, NGC~5253, NGC~4449, NGC~4214, and NGC~625 are known to host young (t$<10$ Myr), massive ($\simeq10^5$ \msun) clusters. In NGC~5253, the dynamical mass of the central star cluster is estimated to be $3\times10^5$ \msun\ \citep{RodriquezRico2007}, while numerous clusters with stellar mass of $10^4$ \msun\ have also been measured \citep{deGrijs2013}. Similarly, in NGC~4449, the central super star cluster has a stellar mass of $2\times10^5$ \msun\ \citep{Boker2001} with up to 20 additional star clusters with masses between 10$^4-10^5$ \msun\ \citep{Annibali2011}. NGC~625 hosts a centralized starburst \citep{Cannon2003}; the H$\alpha$ luminosity of the central cluster approaches that of 30 Doradus \citep{Skillman2003} whose stellar mass is estimated to be $4\times10^5$ \msun\ \citep{Bosch2009}. NGC~4214 includes multiple massive clusters in the central region with masses ranging from $2\times10^4 - 5\times10^5$ \msun\ \citep{Sollima2013}. 

Visual inspection of the HST optical images for these four galaxies used to derive the CMD-based SFRs confirms knots of unresolved stars in these massive clusters. In unresolved clusters, stars that are not photometrically recovered because of blending do not factor into the CMD-based SFRs, but certainly contribute to the integrated UV emission measured in the GALEX images. Therefore, a comparison between CMD-based SFRs and FUV-SFRs will be biased. We include these outliers in the rest of the study for completeness, differentiating them from the rest of the sample with unique plot symbols. However, we do not include them in the calculation of subsequently reported values or in the SFR calibration. 
 
\subsection{Predicting Integrated UV Fluxes from CMD-based SFRs\label{predict}}
The SFRs(t,Z) derived from the resolved stellar populations can be used to model the present-day extinction-free UV luminosity expected from a galaxy, taking into account both the changes in SFR over time and the metallicity of each system. We used four models with different approaches to predict the UV luminosities from the SFHs: (1) the synthetic spectral model $\textsc{P\'{E}GASE 2.0}$\footnote{$\textsc{P\'{E}GASE 2.0}$ is publically available via http://www2.iap.fr/users/fioc/PEGASE.html} \citep{Fioc1997}; (2) synthetic stellar populations generated from {\tt MATCH} \citep{Dolphin2002} which stochastically populates a synthetic UV CMD based on an IMF and the input SFR and metallicity evolution; (3) the synthetic stellar population model Stochastically Lighting Up Galaxies \citep[SLUG;][]{Fumagalli2011,daSilva2012}, which stochastically populates the IMF and includes additional temporal stochastic effects due to clustered star formation; and (4) the stellar population synthesis model GALAXEV \citep{Bruzual2003}. 

\subsubsection{Predicting UV Fluxes Using the Synthetic Spectral Model $\textsc{P\'{e}gase}$}
For the first method, we used the synthetic spectral model  $\textsc{P\'{e}gase}$ 2.0 to generate a full spectrum for each galaxy from our SFHs. The $\textsc{P\'{e}gase}$ code builds a synthetic galaxy spectrum from a set of global parameters such as an initial gas mass, modeled stellar feedback processes, chemical enrichment history, and SFH. The model assumes a fully populated IMF and scales the spectrum by the overall mass of a galaxy. We selected a Salpeter IMF with the mass limits of 0.1$-$120 \msun, to match that used in deriving the SFHs. The use of a single power-law IMF is appropriate as the the stars of interest (both in generating the recent CMD-based SFRs and those generating the UV flux) are higher mass stars not impacted by a double power-law IMF. The $\textsc{P\'{e}gase}$ model includes stellar evolutionary tracks from the main sequence to the white dwarf stage. The older Padova stellar evolutionary isochrones are used \citep{Bertelli1994} without the updates from \citet{Marigo2008}. Dust extinction is modeled using a radiative transfer code. The main input to the $\textsc{P\'{e}gase}$ model is the SFR(t,Z) normalized by the total baryonic mass of the system. The baryonic mass normalization factor was estimated in two steps. First, we estimated the total stellar mass created in the field of view of the HST observations in each galaxy by integrating the lifetime SFHs over all time intervals. Second, the total gas mass was estimated to be approximately equivalent to the stellar mass, as gas-rich dwarf irregulars typically have gas mass fractions of $\sim0.5$ \citep{Begum2008}. Therefore, the normalization factor of total baryonic mass was $2\times \Sigma$~(SFR($\Delta$t)$\times\Delta$t). 

\begin{figure}
\includegraphics[width=0.48\textwidth]{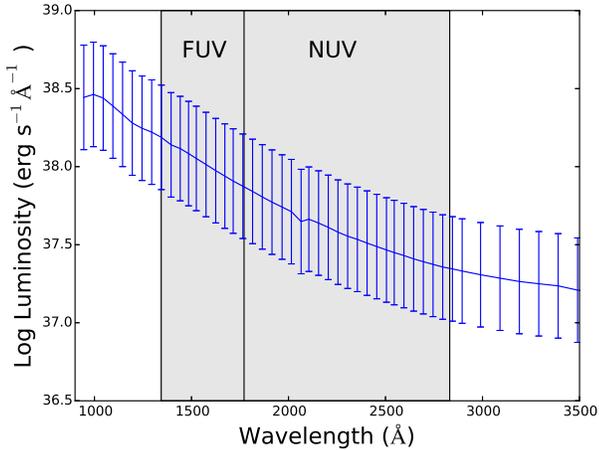}
\caption{Synthetic spectrum of NGC~4068 generated using the $\textsc{P\'{e}gase}$ model and the SFR(t,Z) derived from optical imaging. The uncertainties are based on Monte Carlo simulations of the SFHs. The NUV and FUV bandpasses are highlighted in grey.}
\label{fig:pegase_spectrum}
\end{figure}

In order to include an estimate of uncertainties, we generated synthetic spectra using 50 Monte Carlo simulations of the SFR(t,Z). The final predicted spectrum is the mean of these 50 spectra and the dispersion on the mean provides an estimation of the uncertainty on the spectral luminosity. The predicted integrated UV fluxes were calculated by summing the $\textsc{P\'{e}gase}$ present day spectrum across the GALEX NUV and FUV bandpasses, taking into account the GALEX sensitivity curves for the UV detectors \citep{Morrissey2005}. Figure~\ref{fig:pegase_spectrum} presents an example of the synthetic spectrum for NGC~4068 with uncertainties based on the standard deviation from Monte Carlo simulations. Both the GALEX NUV and FUV bandpasses are highlighted in grey. 

\subsubsection{Predicting UV Fluxes Using Synthetic Stellar Populations from {\tt MATCH}\label{predict_match}}
For the second method, we generated synthetic stellar populations based on the SFRs(t,Z) using the {\tt fake} routine from the CMD fitting code {\tt MATCH} and Padova stellar evolutionary models \citep{Marigo2008}. The stellar populations were generated assuming the distances listed in Table~\ref{tab:galaxies}, and a Salpeter IMF with the same mass limits used in deriving the SFHs. The IMF is randomly populated for the input SFR(t). This is an important stochastic effect to consider at low-SFRs as the upper-end of an IMF may not be fully populated, and, thus, a star forming region may be under-luminous for a given stellar mass (or SFR). Additionally, internal extinction was set to zero in the model. The output from {\tt MATCH} is a list of synthetic magnitudes based on the Vega magnitude system with luminosities in the HST F555W or F606W (depending on the filter used in the actual HST observations), and F814W filters, and the GALEX NUV and FUV filters, using the calibration of the isochrones for the GALEX bandpasses. The UV luminosities of the synthetic stellar populations were converted from Vega magnitudes to the AB magnitude system using the conversions for the GALEX bandpasses from the database in \citet{Marigo2008}:

\begin{equation}
NUV_{AB} = NUV_{Vega} + 1.662
\label{nuv_ab_mag}
\end{equation}
\begin{equation}
FUV_{AB} = FUV_{Vega} + 2.128
\label{fuv_ab_mag}
\end{equation}

\begin{figure}
\includegraphics[width=0.48\textwidth]{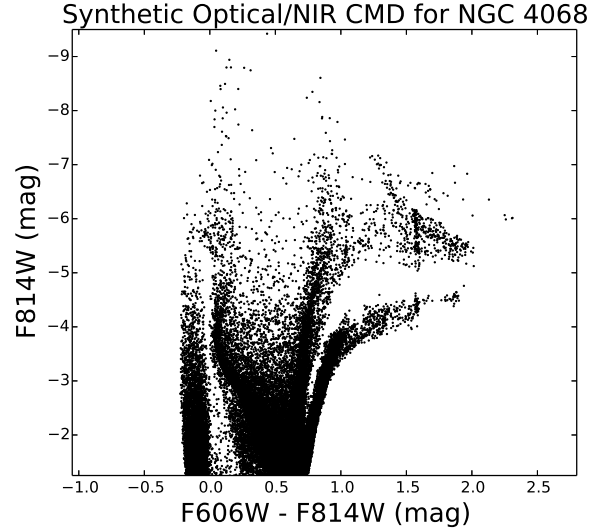}
\caption{Synthetic CMD in the HST filters for NGC~4068 based on the SFH reconstructed from the optical imaging. The main difference between this synthetic CMD and the observed CMD shown in Figure~\ref{fig:hst_cmd} is the absence of observational errors in the synthetic stellar population. While it is possible to include errors using artificial star tests, this introduces additional uncertainties when predicting the UV flux expected from the CMD.}
\label{fig:compare_cmds}
\end{figure}

\begin{figure}
\includegraphics[width=0.48\textwidth]{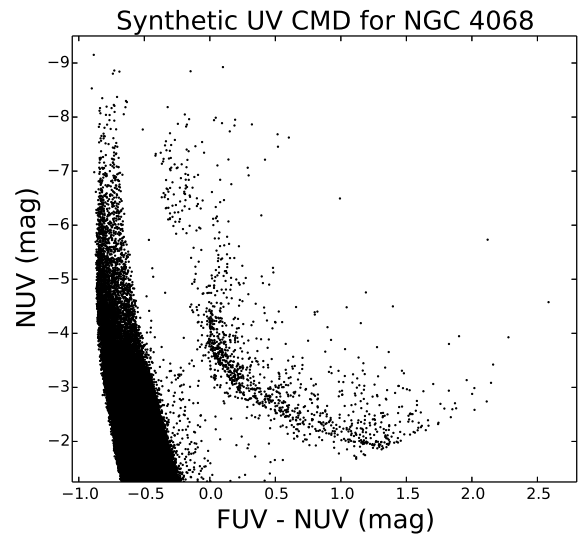}
\caption{Synthetic CMD in the GALEX filters for NGC~4068 based on the SFH reconstructed from the optical imaging. This CMD is the UV companion to the synthetic CMD in the HST filters shown in Figure~\ref{fig:compare_cmds}. The majority of the UV emission is produced by the upper MS and blue HeB stars.}
\label{fig:galex_cmd}
\end{figure}

\noindent As a consistency check, synthetic CMDs generated in the HST filters can be compared with the observed CMDs. Figure~\ref{fig:compare_cmds} shows the synthetic CMD generated for NGC~4068 from the SFH, which can be compared with observed CMD shown in Figure~\ref{fig:hst_cmd}. The main difference between the two CMDs is due to the generation of the synthetic stellar populations without observational errors. Figure~\ref{fig:galex_cmd} shows the corresponding synthetic CMD for NGC~4068 in the GALEX filters. The predicted integrated UV luminosities were determined from the synthetic stellar populations by summing the UV luminosities in the GALEX filters. 

\subsubsection{Predicting UV Fluxes Using Synthetic Photometry from SLUG\label{predict_slug}}
For the third method, we generated integrated UV fluxes based on the SFRs(t) at the present day metallicity using SLUG. SLUG is a modification of the Starburst99 code which uses Monte Carlo realizations to account for the sampling stochasticity of the upper-end of the IMF, temporal stochasticity from the finite sampling of the SFH, and stochastic sampling of the initial cluster mass function (ICMF). The SLUG model creates and evolves a synthetic galaxy over a maximum of 1 Gyr and generates the expected flux at different bandpasses, including the GALEX NUV and FUV bands. For consistency, the Padova stellar evolutionary models of \citet{Marigo2008} were used and the same IMF slope was assumed but with mass limits between 0.1 and 100 \msun. However, in contrast to the other models, the galaxies are modeled using values of their present day metallicity and no chemical evolution is included. This will not have a significant effect on our comparison as there is little chemical evolution expected over the last 1 Gyr. However, the metallicity inputs for SLUG are approximate as the choice of isochrones is limited to 0.0004, 0.004, and 0.008 without any intermediary interpolation. This will have some effect on the predicted FUV fluxes as the FUV emission from stars varies with metallicity \citep[e.g.,][]{Madau2014}. We discuss the dependency of FUV emission on metallicity further in Section~7.

The SLUG model was developed to probe the effects of stochastic variations that can occur in stellar populations at low absolute SFRs. Similar to {\tt MATCH}, SLUG randomly populates the upper-end of an IMF for a given input SFR. Additionally, SLUG can be fine-tuned to allow for unclustered or clustered star formation. In this context, a `cluster' is meant to include any type of stellar association, dynamically bound or not. For a clustered star formation scenario, SLUG randomly populates each individual cluster with the assumed IMF. Generally, this results in lower UV fluxes for clustered SF, particularly if the cluster masses are constrained to be low-mass clusters, as UV-bright, high-mass stars have a lower probability of occurring in low-mass clusters.   

\subsubsection{Predicting UV Fluxes Using Stellar Population Synthesis Models from GALAXEV\label{predict_galaxev}}
For the fourth and final method, we used the stellar population synthesis model GALAXEV to generate a full spectrum for each galaxy from our CMD-based SFHs. Similar to $\textsc{P\'{e}gase}$, the GALAXEV code builds a synthetic spectrum of a galaxy based on the SFH but assumes a constant metallicity over the lifetime of the system. This will not have significant effect on our comparison as there is little chemical evolution expected over the last 1 Gyr. However, similarly to the SLUG model, the metallicity inputs are approximate as the choice of isochrones is limited to 0.0004, 0.004, and 0.008 without any intermediary interpolation. For consistency with the other predictions, we assumed the same IMF slope. However, the mass limits for a Salpeter IMF in GALAXEV is set from 0.1 and 100 \msun (see below for a discussion on the impact of the IMF mass limits). In addition, the Padova stellar evolutionary models \citep{Marigo2008} were used. From \citet{Bruzual2003}, the older 1994 Padova models are preferred over these because of differences in temperatures of the red giant branch stars. However, not only do these differences not impact the UV fluxes, but constancy over the choice of stellar evolution models in the predictions is preferred. No dust attenuation was included in the spectra generated using GALAXEV.

\subsubsection{Comparison of Inputs and Assumptions Between the Four Models\label{compare_inputs}}
Wherever possible, we used the same inputs and basic assumptions in all four models to predict the UV fluxes. As described above, all the models use the CMD-based SFRs(t) measured from the resolved stellar populations and the Padova stellar evolution libraries. Further, no reddening corrections were applied making the UV luminosities all intrinsic, dust-free luminosities. However, there are some differences between the models. The most important difference is in the choice of stellar atmosphere models used to transform each model's results into a spectrum or flux measurement. We discuss these differences in detail below (Section~4). 

There are also smaller differences between in the models. For example, {\tt MATCH} applies a user-specified binary fraction of 35\% to generate binary stars in the synthetic populations based on a flat secondary distribution. In contrast, $\textsc{P\'{e}gase}$, SLUG, and GALAXEV do not consider the contribution of binaries to the overall galaxy flux or spectrum. A Salpeter IMF with mass limits from 0.1 to 120 \msun\ was assumed in the $\textsc{P\'{e}gase}$ and {\tt MATCH}  models; SLUG and GALAXEV used a slightly smaller range of 0.1 to 100 \msun. This will have a negligible effect on our results as few, if any, of these higher mass stars are expected at these low SFRs. We found that changing the IMF mass limits to 0.1$-$100 \msun\ had less than a 5\% increase in the predicted fluxes from the synthetic spectral model $\textsc{P\'{e}gase}$. Both $\textsc{P\'{e}gase}$ and {\tt MATCH} utilize the chemical enrichment history information derived from the stellar populations and interpolate between isochrones, while SLUG and GALAXEV assume a constant metallicity over the past 1 Gyr approximated at one of three isochrones. As noted above, the enrichment history will likely have a negligible effect on our results as there is little expected chemical evolution in the last 1 Gyr, but small differences can be introduced when using approximate metallicities. 

As mentioned above, the SLUG model can be altered from populating the IMF in an unclustered star formation scenario to one with clustered star formation. We found the UV flux predictions from SLUG depended on whether the star formation was clustered or not, but were not as sensitive to the degree of clustering (i.e., there was an appreciable effect on the flux predictions between assuming $\textit{no}$ stars formed in clusters versus 100\% of the stars formed in clusters, but there were only small differences between assuming 50\%, 70\% or 100\% of the stars formed in clusters). Additionally, the SLUG predictions depended on the range of assumed cluster masses. 

\subsection{Calculating Extinction Corrected UV Luminosities\label{uv_lum}}
The measured NUV and FUV fluxes in Table~\ref{tab:flux_lum} were converted to luminosities in the AB magnitude system using the calibration and zeropoint magnitudes \citep{Morrissey2007}$\colon$
\begin{align}
& NUV = -2.5 \cdot \textrm{log}_{10}~(\textrm{Flux}_\textrm{NUV} \nonumber \\
& /~2.06 \times 10^{-16}~ \textrm{erg~sec}^{-1}~\textrm{cm}^{-2}~\textrm{\AA}^{-1}\textrm{)}~+~20.08\label{eq:nuv_lum}
\end{align}
\begin{align}
& FUV = -2.5 \cdot \textrm{log}_{10}~(\textrm{Flux}_\textrm{FUV} \nonumber \\
& /~1.40 \times 10^{-15}~ \textrm{erg~sec}^{-1}~\textrm{cm}^{-2}~\textrm{\AA}^{-1}\textrm{)}~+~18.82\label{eq:fuv_lum}
\end{align}

\noindent These UV magnitudes are based solely on the UV flux measured in each galaxy and do not account for possible UV extinction along the line of sight. Therefore, we estimated the attenuation expected at UV wavelengths. Although many of the galaxies in our sample are low metallicity dwarfs, and thus are expected to have low internal dust content and correspondingly low values of UV extinction, eight systems have somewhat higher metallicities (i.e., $8.1 <$ log 12 $+$ (O/H) $<8.4$; see Table~\ref{tab:galaxies}). 

One way to estimate the amount of attenuation at UV wavelengths is to measure the total IR emission from the dust, which is primarily heated by UV emission from stars, and convert this IR flux to an equivalent UV luminosity. This method is an improvement over using the FUV-NUV color to estimate extinction, which relies on assumptions about the intrinsic color of the emitting stellar populations, dust extinction, geometry of the system, and the effects of scattering \citep[e.g.,][]{Gordon2001, Misselt2001}. The relationship between attenuation of UV fluxes and mid-IR emission from dust was derived empirically by \citet{BuatXu1996} for a range of extinction mechanisms. More recently, the attenuation expected at the GALEX bandpasses extrapolated from a measurement of the total IR flux was derived using models of active star forming regions \citep{Buat2005}. While this prescription cannot differentiate between IR flux from dust heated by UV emitting stars and dust heated by stars emitting longwards of the UV, differences in results from more massive starburst galaxies with larger populations of UV emitting stars and from normal star-forming galaxies are only of order 30\% \citep[e.g.,][]{Buat1999, Meurer1999, Gordon2000, Hao2011}. We use the empirical relations from \citet{Buat2005} derived from a sample of spiral and irregular galaxies that extend to an M$_{NUV} = 16$ mag (AB system), the approximate lower limit of our sample. We reproduce the relations here in Equations~\ref{equation:a_fuv} and \ref{equation:a_nuv}:
\begin{align}
& A_{\textrm{NUV}} = -0.0495\cdot x^{3} + 0.4718\cdot x^{2} + 0.8998\cdot x + 0.2269, \nonumber\\
& \textrm{where x}=\textrm{log(F}_\textrm{TIR}\textrm{/F}_\textrm{NUV}) \label{equation:a_fuv}
\end{align}

\begin{align}
& A_{\textrm{FUV}} = -0.0333\cdot y^{3} + 0.3522\cdot y^{2} + 1.1960\cdot y + 0.4967, \nonumber\\
& \textrm{where y}=\textrm{log(F}_\textrm{TIR}\textrm{/F}_\textrm{FUV}) \label{equation:a_nuv}
\end{align}

\noindent The total IR flux needed for the above equations can be estimated using the MIPS flux measurements following the empirical prescription from \citet{Dale2002}:
\begin{align}
& \textrm{F}_\textrm{TIR} = 1.559\cdot\nu\cdot\textit{f}_{\nu}(24 \micron) + 0.7686\cdot\nu\cdot\textit{f}_{\nu}(70 \micron)\nonumber \\
& + 1.347\cdot\nu\cdot\textit{f}_{\nu}(160 \micron)
\label{equation:ftir}
\end{align}
\noindent The MIPS fluxes in Table~\ref{tab:flux_lum} and the relationships in Equations~\ref{equation:a_fuv}-\ref{equation:ftir} were used to correct the NUV and FUV luminosities for extinction. Table~\ref{tab:lum_sfrs} lists the extinction estimates (A$_{\rm{NUV}}$ (mag) and A$_{\rm{FUV}}$ (mag)), and the extinction corrected NUV and FUV luminosities. For completeness, this correction was performed for the entire sample, regardless of metallicity. Two galaxies, Antlia and UGC~9128, are non-detections in all three MIPS images; the UV luminosities were not adjusted for these systems. Four galaxies (UGC~6456, IC~4662, NGC~1569, and NGC~6789) were not observed in all three MIPS channels; their estimated extinction values are lower limits as noted in Table~\ref{tab:lum_sfrs}. 

\subsection{Calculating SFRs from Existing Scaling Relations\label{uv_sfrs}}
As discussed in Section~1, there are a number of theoretical scaling relations to convert UV luminosities to SFRs. Both the earlier calibration of the FUV-based SFR scaling relation by \citet{Kennicutt1998} and the recent updated calibrations from both \citet{Hao2011} and \citet{Murphy2011} are based on the Starburst99 spectral models \citep{Leitherer1999, Vazquez2005}. The Starburst99 code uses an isochrone synthesis technique to generate a synthetic spectrum for a galaxy based on a given set of user-specified parameters (for example, a constant SFR and constant metallicity over some time period). Briefly, Starburst99 uses a stellar evolutionary library and an assumed IMF to generate a Hertzprung-Russell diagram, and then applies stellar atmospheric models to points in the binned Hertzprung-Russell diagram \citep{Leitherer1995}. The cumulative sum of the modeled spectra results in a synthetic galaxy spectrum. The monochromatic flux in the synthetic spectrum at an FUV wavelength (generally taken as 1500 \AA) is then used as a basis to calibrate the theoretical integrated FUV-based SFR scaling relation. 

We chose to use the following scaling relation from \citet{Hao2011} based on the updated models from Starburst99 to convert our extinction corrected, integrated UV luminosities into SFRs:
  \begin{equation}
     \textrm{SFR} = 1.33\times10^{-28}  \times L^0_{\nu},
     \label{eq:uv_sfr}
  \end{equation}
where SFR is in $M_{\Sun}$ yr$^{-1}$ and L$^0_{\nu}$ is the attenuation corrected FUV luminosity in ergs s$^{-1}$ Hz$^{-1}$. The FUV-based SFRs calculated in this way are listed in Table~\ref{tab:lum_sfrs}. The conversion factor in Equation~\ref{eq:uv_sfr} was established specifically for the GALEX FUV bandpass and assumes a Salpeter IMF with mass limits of $0.1-100$ \msun. The upper mass limit of 100 \msun\ is lower than the 120 \msun\ limit used in deriving the SFHs and in two of the models used for predicting the UV fluxes. We found through experimenting with different mass limits in the synthetic spectral model $\textsc{P\'{e}gase}$ that this upper limit has $<5$\% effect on the results (see discussion above).  The SFRs using the scaling relation from \citet{Hao2011} are $\sim$5\% lower than those calculated from the theoretical scaling relation by \citet{Kennicutt1998} and 23\% higher than the SFRs calculated using the empirical relation from \citet{Salim2007}. However, as these scaling relations differ by a linear offset, it is simple to consider the effect of this choice on our results (see Section~5).

\section{Comparison of the Measured and Predicted UV Fluxes \label{compare_uv_integrated}}
One of the critical tests to our understanding of star formation measurements is whether the measurements across wavelength regimes agree. For this study, the different assumptions used in determining the CMD-based and FUV-based SFRs $and$ the calibration of the FUV scaling relation are embedded in the comparison. In this section, we simplify the comparison by eliminating the calibration of the FUV-based SFR. We compare only the UV fluxes modeled from the independently measured CMD-based SFRs with the UV fluxes  measured from GALEX images. This test therefore integrates the resolved stellar population measurements with the synthetic spectral models and stellar population synthesis models, without the assumptions in the FUV-SFR scaling relation. In the following section,  we will expand our analysis and examine the SFR values from both the CMD analysis and the FUV scaling relations.

\subsection{NUV Ratios\label{nuv_ratios}}
\begin{figure}
\includegraphics[width=0.5\textwidth]{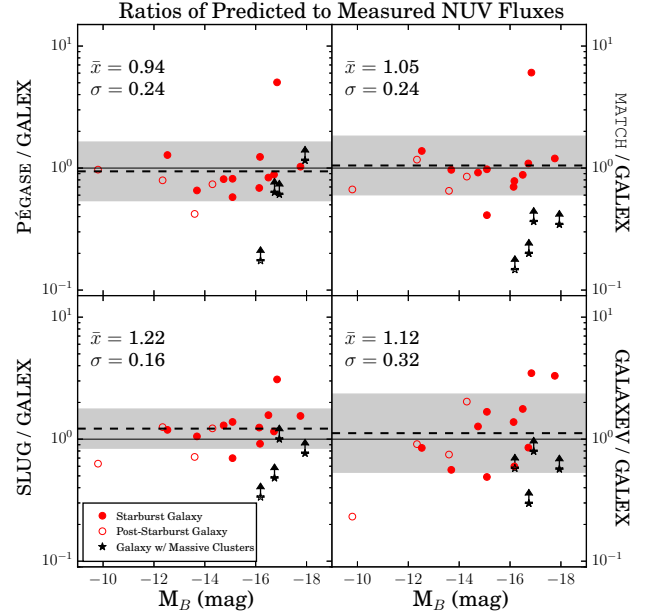}
\caption{Ratios of the predicted to observed NUV fluxes as a function of the B-band absolute magnitude. The predictions were generated using the CMD-based SFRs and the synthetic spectral model (top left), synthetic stellar populations (top right), a synthetic galaxy taking into account stochastic effects (bottom left), and a synthetic galaxy spectrum (bottom right). Starburst galaxies are plotted as filled symbols; post-starburst galaxies are plotted with unfilled symbols. The dashed line represents the mean NUV ratio and the shaded grey regions denote the standard deviation on the mean. The galaxies hosting young, massive clusters are plotted as stars and are excluded from the calculation of the mean ratio.}
\label{fig:nuv_galex_models}
\end{figure}

The ratios of the measured to predicted NUV fluxes from the four models are reported in Table~\ref{tab:galex_models} and shown as a function of B-band luminosities in Figure~\ref{fig:nuv_galex_models}. The FUV ratios are shown in Figure~\ref{fig:fuv_galex_models} and are discussed separately in Section~4.2. Each panel in Figure~\ref{fig:nuv_galex_models} presents the results from one of the four models as labelled. Starburst galaxies are plotted as filled symbols and post-starburst galaxies are plotted as unfilled symbols. The thin solid lines represent a perfect match between the predicted and measured values, the dashed lines represent the mean value assuming a log-normal distribution, and the shaded regions represent the standard deviation on the log-normal mean. For completeness, we include the four galaxies whose CMD-based SFRs were previously identified as lower-limits (plotted as stars), but do not include their NUV flux ratios in the mean. As expected, the predictions for these galaxies are biased low.

From Figure~\ref{fig:nuv_galex_models}, the NUV fluxes predicted by all four models agree with the measured fluxes for the majority of the sample. This is remarkable given the complex inputs inherent in this comparison. The mean ratios  and standard deviations based on a log-normal distribution range from $\bar{x} = 0.94, \sigma=0.24$ to  $\bar{x} = 1.22, \sigma=0.16$ as noted in each panel. The generally high dispersion is not unexpected. The modeled fluxes are based on the SFRs(t) measured from the resolved stellar populations. However, even with the fine temporal resolution achievable from the STARBIRDS data on these galaxies at recent times, the SFRs(t) are not $\textit{continuously}$ sampled. Fluctuations in the SFR are known to occur in the sample on timescales shorter than a few Myr \citep[e.g.,][]{McQuinn2010b}. This temporal stochasticity can occur more frequently at lower SFRs in low-mass galaxies as the SFRs are measured over a small number of individual star-forming events. The standard deviations between the predicted and measured flux values are consistent with the types of variations one would expect given the finite temporal sampling of the input SFHs. Our result is in general agreement with modeling of the spectral energy distributions of dwarf galaxies that show changes in the SFR on short timescales can account for discrepancies in the flux predictions of up to a factor of 2 \citep{Johnson2013}. The flux for one of the higher luminosity galaxies, NGC~6822, is consistently over-predicted by the models. This system is unique in the STARBIRDS sample as the star-formation activity has increased significantly over the most recent time bin of 20 Myr. It is possible that the SFR has fluctuated significantly during this shorter time period, which would affect the flux predictions.

The overall agreement between the NUV flux predictions from the four different models and the measured fluxes provides an independent check on a number of variables. First, the UV attenuation has been adequately corrected in the GALEX fluxes using the infrared flux from the MIPS images. If the measured fluxes were significantly affected by extinction at UV wavelengths, the predictions would be systematically biased high. Second, the SFHs based on CMD fitting are accurately measuring the SF at recent times in these systems. Third, the transformations from the stellar evolution libraries to NUV wavelengths are consistent with what is measured from the composite stellar populations. Fourth, stochastic sampling of the IMF at lower SFRs does not appear to be a significant factor in our results as the predictions from both $\textsc{P\'{e}gase}$ and GALAXEV, which do not include stochastic effects, are in agreement with the measured values and consistent with predictions from {\tt MATCH} and SLUG, which do include include such effects. This is not surprising as the stellar populations responsible for producing the majority of the NUV fluxes includes the comparatively more abundant stars of lower masses than the FUV emission, reducing the impact of whether or not the upper end of the IMF was fully populated.

\subsection{FUV Ratios\label{fuv_ratios}}
\begin{figure}
\includegraphics[width=0.5\textwidth]{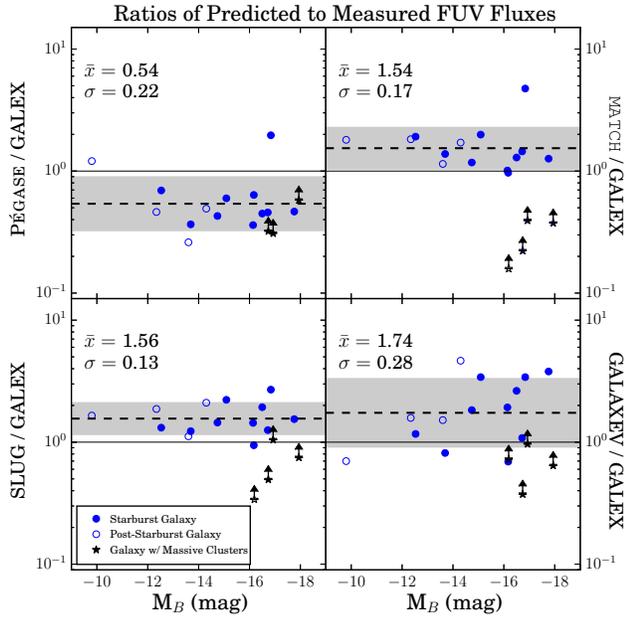}
\caption{Ratios of the predicted to observed FUV fluxes as a function of the B-band absolute magnitude. The predictions were generated using the CMD-based SFRs and the synthetic spectral model (top left), synthetic stellar populations (top right), a synthetic galaxy taking into account stochastic effects (bottom left), and a synthetic galaxy spectrum (bottom right). Starburst galaxies are plotted as filled symbols; post-starburst galaxies are plotted with unfilled symbols. The dashed line represents the mean FUV ratio and the shaded grey regions denote the standard deviation on the mean. The galaxies hosting young, massive clusters are plotted as stars and are excluded from the calculation of the mean ratio.}
\label{fig:fuv_galex_models}
\end{figure}

Figure~\ref{fig:fuv_galex_models} shows a similar analysis to Figure~\ref{fig:nuv_galex_models} for the FUV bandpass. Here, the predicted FUV fluxes from the four models are seen to disagree with the measured GALEX FUV fluxes. The mean predicted to measured FUV ratio for $\textsc{P\'{e}gase}$ is low with $\bar{x} = 0.54, \sigma=0.22$, whereas the mean FUV ratios for {\tt MATCH}, SLUG, and GALAXEV are consistent with one another but high ranging from $\bar{x} = 1.54, \sigma=0.17$ to $\bar{x} = 1.56, \sigma=0.13$  and $\bar{x} = 1.74, \sigma=0.28$. The SLUG results show a lower dispersion than the other models. This is most likely due to the inclusion of additional stochastic effects when clustered star formation is taken into account. 

The offsets in the predicted FUV fluxes are likely due to the calibration of the models at the FUV bandpass, which depends on both the application of stellar atmosphere libraries and the calibration of the stellar evolutionary isochrones at FUV wavelengths. As noted in Section \ref{compare_inputs}, there are important differences between the four models that impact the calibrations at FUV wavelengths. All four codes use different stellar atmosphere models in the UV regime as summarized in Table~\ref{tab:stellar_atm_models}. $\textsc{P\'{e}gase}$, SLUG, GALAXEV use the models from \citet{Lejeune1997, Lejeune1998} (based on the original grids of model atmosphere spectra from \citet{Bessell1989}, \citet{Fluks1994}, and \citet{Kurucz1992}) for the majority of stars contributing to the galaxy spectra. However, SLUG and GALAXEV use additional models for stars with strong stellar winds, O stars, and non-solar metallicity stars. {\tt MATCH} uses the library from \citet{Castelli2004}, that has some overlap with the original \citet{Kurucz1992} model spectra. In addition, {\tt MATCH} uses a zero point transformation for the isochrones to the GALEX FUV bandpass. 

While disentangling the different effects of these libraries and zero point calibrations is beyond the scope of this study, we did perform a series of simple tests to measure the impact stellar atmosphere models can have on FUV fluxes. We ran a series of simulations using the Starburst99 spectral models, which allows the user to model a galaxy spectrum with different atmospheric libraries. We used the same stellar evolutionary libraries and IMF as our FUV predictions, keeping the inputs and assumptions as closely matched as possible. The simulations spanned a SFR range from 0.001 to 1.0 \msun\ yr$^{-1}$ and a metallicity range from $Z$ of 0.0004 to 0.02. While the use of another code may introduce additional uncertainty in the comparison, the Starburst99 code provides a relative measure of how the stellar atmosphere models affect the FUV fluxes. Further, not only is the technique used in Starburst99 similar to GALAXEV, the SLUG code is based on the Starburst99 code, with modifications to include stochastic effects. Thus, the additional uncertainties in the relative comparison should be small.

Starburst99 simulations were run using the atmospheres applied in SLUG and the atmospheres used in $\textsc{P\'{e}gase}$ with the matched stellar evolution isochrones \citep{Bressan1993}; the atmospheres used in {\tt MATCH} and GALAXEV were not available in the Starburst99 simulations. The Starburst99 fluxes using the \citet{Lejeune1997, Lejeune1998} and  \citet{Hillier1998} models, similar to the SLUG stellar atmosphere models, varied by up to 10\% at solar metallicities, with smaller differences at the lower metallicities of the STARBIRDS sample. The Starburst99 fluxes using just the \citet{Lejeune1997, Lejeune1998} models with the older Padova models, similar to $\textsc{P\'{e}gase}$, varied by a similar amount with a similar dependency on metallicity. 

While our simple test using Starburst99 did not produce big enough changes in the FUV fluxes to account for the discrepancies seen in Figure~\ref{fig:fuv_galex_models}, we cannot rule out differences in the application of the atmospheres as a significant source of the offset. \citet{Bruzual2003} note in their Appendix that the UV colors of the spectra are significantly more uncertain than the optical colors due to the lack of comparison standards at non-solar metallicities \citep{Westera2002}. They qualitatively conclude that the reliability of the predictions for young stars at low metallicity is fair to poor. A previous comparison between $\textsc{P\'{e}gase}$  and Starburst99 reported significant differences \citep{Kewley2001}. However, not only were the model atmospheres different, but the earlier version of the Starburst99 code used a different stellar evolutionary library. Thus, while \citet{Kewley2001} speculated that the different model atmospheres were driving the differences, the different evolutionary libraries prohibited isolating the effects of just the model atmospheres. 

\section{Comparison of SFRs from Integrated FUV Emission using Existing Scaling Relations with SFRs(t) from Resolved Stellar Populations \label{compare_sfrs}}
\begin{figure}
\includegraphics[width=0.5\textwidth]{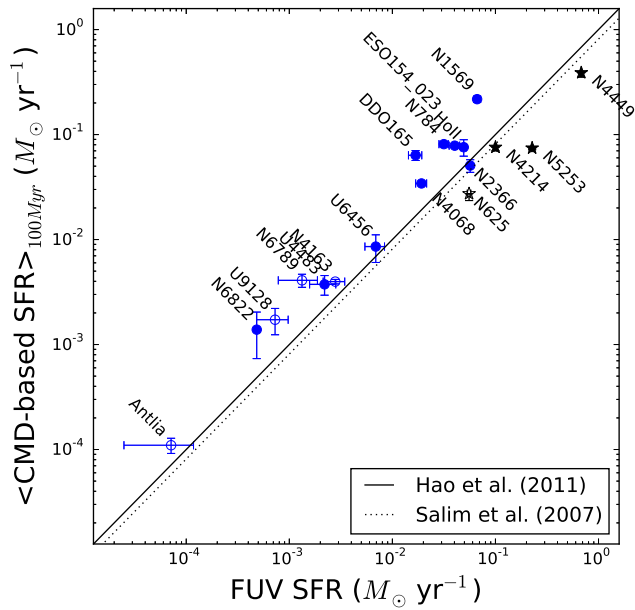}
\caption{Comparison of the CMD-based SFRs averaged over the last 100 Myr with FUV-based SFRs calculated using the scaling relation from \citet{Hao2011}. The uncertainties on the CMD-based SFRs are based on the standard deviation in SFRs averaged over 50, 100, and 150 Myr, thus providing a measure of how variable the SFRs have been over the timescale of interest. Starburst galaxies are plotted as filled circles; post-starburst galaxies are plotted as unfilled circles. The solid black line indicates a ratio of unity between the two SFRs; the dotted line is where agreement would be using the FUV-based SFRs scaling relation from \citet{Salim2007}. The CMD-based SFRs are higher than the FUV-based SFRs for all systems except in four cases, plotted as stars, whose CMD-based SFRs are lower-limits and which also compromise the assumption of constant SF at the most recent times as discussed in the text.}
\label{fig:opt_uv_sfrs}
\end{figure}

Figure~\ref{fig:opt_uv_sfrs} presents a comparison of the SFRs measured from the resolved stellar populations averaged over the past 100 Myr to the SFRs based on the extinction corrected, integrated FUV emission using the scaling relation from \citet{Hao2011}. As noted above, the uncertainties on the CMD-based SFRs are the standard deviations of the SFRs between timescales of 50, 100, and 150 Myr. We included the time bin of 150~Myr in the standard deviation as it extends the temporal baseline over which assumption of constant SFR can be evaluated. The uncertainties in the FUV-based SFRs include statistical uncertainties only. For completeness, we include the four galaxies with young, massive clusters whose CMD-based SFRS are lower-limits (plotted as stars). As expected, these galaxies have higher FUV-based SFRs than the CMD-based SFRs. 

The FUV-based SFRs are clearly correlated with CMD-based SFRs. However, the majority of the FUV-based SFRs are offset to lower values by up to a factor of 4. The comparison to the FUV-based SFRs is affected by our choice of the scaling relation we used to convert from the FUV emission to a SFR (see \S\ref{uv_sfrs}). If we had used the relation from \citet{Salim2007}, the FUV-based SFRs would be 25\% lower, represented by the dotted line in Figure~\ref{fig:opt_uv_sfrs}. This difference would increase the discrepancies between the CMD-based and FUV-SFRs. Further, in deriving the empirical scaling relation to convert the FUV luminosity to a SFR, \citet{Salim2007} calibrated the results using SFRs derived from fitting spectral energy distributions from the UV to optical bands. In their work, the empirical SFR relation is noted to be a poorer fit to the SED-derived SFRs for values below $\sim$0.1 \msun\ yr$^{-1}$; the empirical relation generally producing FUV based SFRs that are $\textit{higher}$ than SED based UV SFRs with a significant amount of scatter. The authors attribute the differences to difficulties in accurately estimating the extinction based on FUV-NUV colors. Thus, not only do our results show $\textit{lower}$ FUV-based SFRs for the majority of the galaxies, but our analysis also corrects for extinction based on the far IR emission from dust in the galaxies. 

On the one hand, SFRs calculated from integrated FUV emission measurements are useful because of their simple application. On the other hand, interpreting the SFRs calculated in this way can be limited for the same reason. Without additional information (that can be difficult to obtain for large surveys or distant galaxies), the sources of ambiguity in the interpretations include correcting for FUV attenuation from dust, evaluating whether the assumption of constant SFR is valid for a given system, and determining whether the data are fully sampling the IMF and/or the SFH of the system. Indeed, many of these variables are often invoked as plausible explanations for discrepancies in observational comparison of integrated light SFRs. While these challenges are present in this study, the advantage of comparing integrated FUV flux measurements with SFHs derived from resolved stellar populations is that a number of the variables can be more fully constrained. In Appendix~A, we discuss several of these factors in detail and provide arguments to rule out these often invoked uncertainties as the source of the offset between the SFR indicators.

The direction and amount of offset between the FUV-based and CMD-based SFRs is consistent with the offset between the FUV flux predictions from three previous models and the extinction corrected FUV emission measured from the GALEX images (see Section~4.2). This is in contrast to the NUV flux predictions using the same inputs which were in very good agreement with the measured NUV emission (see Section~4.1). Two of the three models, GALAXEV and SLUG, have previously been shown to be consistent with Starburst99 model \citep{daSilva2012} which is used to calibrate the FUV-SFR scaling relation from \citet{Hao2011}. This strongly suggests that the most likely cause of the offset between both the two SFR indicators and the predicted to measured FUV fluxes is the calibration of the FUV flux used in the models, either from the rather uncertain stellar atmosphere models for UV producing stars, a systematic in the stellar evolutionary isochrones at FUV wavelengths, or both. The lower FUV-based SFR for a given CMD-based SFR implies that the theoretical models used in Starburst99 to calibrate the FUV-SFR scaling relation over-predict the FUV flux for a given SFR.

\section{Empirical Calibration of the FUV-based SFR Relation using CMD-based SFRs\label{calibration}}
Here, we quantify the difference between the CMD-based SFRs and the FUV-based SFRs for the STARBIRDS sample. We present an alternative calibration for converting integrated FUV emission to a SFR using the galaxies from the STARBIRDS sample that best meet the assumption of constant SF activity over the lifetime of the FUV emitting stellar population. First, we eliminate the four galaxies that host young stellar clusters with masses of order $10^5$ \msun\ that both compromise the assumption of constant SF and have unresolved stars in the clusters in the optical images (i.e., NGC~4449, NGC~5253, NGC~4214, and NGC~625). Second, we eliminate three galaxies that show significant variability in their CMD-based SFHs over the past 100 Myr, namely Antlia, UGC~9128, and the newly bursting galaxy NGC~6822. 

We adopt the notation for the FUV-based SFR relation introduced by \citet{Charlot2001} where the luminosity of a galaxy is related to a SFR by the efficiency factor $\eta$:

\begin{equation}
\eta_{FUV}^0 = L_{FUV}^0/\langle \textrm{CMD-based}~SFR \rangle_{100~\textrm{Myr}}
\end{equation}

\begin{figure}
\includegraphics[width=0.5\textwidth]{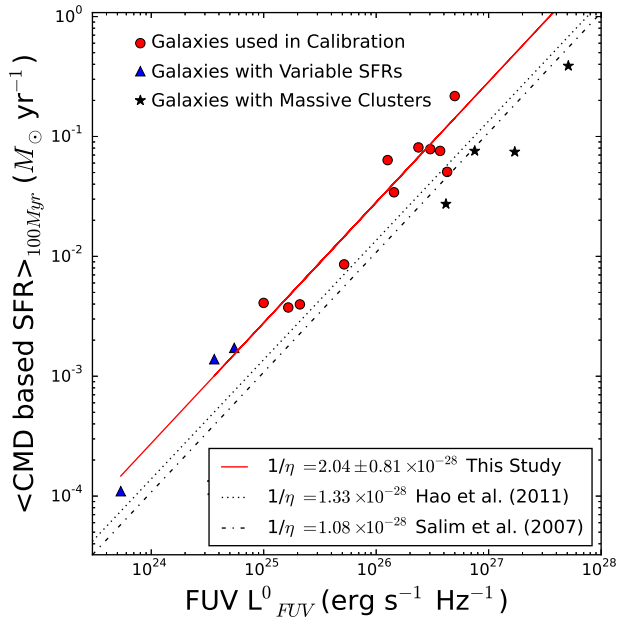}
\caption{The CMD-based SFRs are compared with the extinction corrected FUV luminosities. The red circles represent the galaxies whose SF activity is fairly constant over 100 Myr; uncertainties in the HST SFRs are approximately the size of the plot points. Individual errors are plotted in Figure~\ref{fig:opt_uv_sfrs}. The least squares fit to these points is shown as a solid red line. For comparison, we plot the calibration from \citet{Salim2007} as a dash-dot line and the calibration from \citet{Hao2011} as a dotted line. The different values of the calibrations are noted in the Figure legend. The black stars represent galaxies with young, massive star clusters whose CMD-based SFRs are under-reported and which also compromise the assumption of constant SF at the most recent times. The blue triangles represent the galaxies that show variations in the SF activity at older times. While these are excluded from the calibration, their SFH variations do not impact their FUV emission as much as in systems with significant variations within the last 10 Myr.}
\label{fig:calibration}
\end{figure}

In Figure~\ref{fig:calibration}, we plot the CMD-based SFRs and the extinction corrected FUV luminosities for the sample. Using the galaxies that best approximate constant SFR over the last 100 Myr (red circles), a least squares fit including uncertainties on the CMD-based SFRs yields a value of log~$\eta_{FUV}^0 = -27.69$. The inverse of $\eta_{FUV}^0$ gives the conversion factor for the FUV-based SFR as:

\begin{equation}
SFR = 2.04\pm0.81 \times 10^{-28} \times L_{FUV}^0
	\label{eq:sfr_calibration}
\end{equation}

\noindent where SFR is in \msun~yr$^{-1}$ and $L_{FUV}^0$ is the extinction corrected luminosity measured in the GALEX FUV bandpass in erg s$^{-1}$ Hz$^{-1}$. This calibration is based on a randomly populated Salpeter IMF with mass limits between 0.1 and 120 \msun, for low-metallicity dwarfs with SFRs between $10^{-3}$ and $10^{-1}$ \msun~yr$^{-1}$. This is 1.53 times greater than the calibration reported in \citet{Hao2011} and 1.72 times greater than the calibration reported in \citet{Salim2007} for higher-metallicity and higher-SFR galaxies. Our empirically determined scaling factor includes an estimate of the uncertainty based on the standard deviation of the data from the best fit line, including uncertainties on the individual measurements. This offers an improvement over theoretical calibrations that do not quantify the uncertainties. Table~\ref{tab:lum_sfrs} lists the FUV SFRs for our sample calculated with the new calibration in Equation~\ref{eq:sfr_calibration}.

Note that our sample is small, and therefore the calibration presented here could be improved upon with a larger number of galaxies. However, even with a larger sample of galaxies, there will likely still be a higher level of dispersion in the calibration in the low-SFR regime of dwarf galaxies. To test whether the dispersion in the correlation between SFR to FUV luminosity is due to observational uncertainties or is intrinsic to the relationship, we fit a linear relationship using the IDL \textsc{mpfitexy} routine \citep[][see Section 4]{Williams2010}. The \textsc{mpfitexy} routine depends on the \textsc{mpfit} package \citet{Markwardt2009}. This procedure finds a high degree of intrinsic scatter (97\%) or $\sigma_{instrinic} = 0.79\times 10^{-28}$ from Equation~\ref{eq:sfr_calibration}. The high intrinsic dispersion reflects the complex nature of these systems, but also note that this value would be lower if the reported uncertainties are underestimated. We discuss the dispersion between the SFR and FUV luminosity further in the next section.

Similar findings have been reported by \citet{Chomiuk2011} in a comparison of SFRs measured in the Milky Way with a variety of SFR indicators. These authors report that Galactic SFRs derived for \HII\ regions using resolved low and intermediate mass stellar populations were systematically higher by factors of $\sim2$ to 3 than SFRs derived using high mass stellar tracers in the mid-IR and radio emission. A separate study reported SFRs derived from H$\beta$ line strengths to be $\sim1.8\times$ higher than those derived from UV fluxes in a sample of galaxies at a redshift of $\sim2$ \citep{Zeimann2014}.

In contrast, comparisons of integrated FUV-based SFRs to SFRs derived from SED fits have found the FUV-based SFRs to yield higher SFRs. \citet{Huang2012} compared integrated FUV-based SFRs with SED fits of a sub-sample of low-mass galaxies from the Arecibo Legacy Fast ALFA (ALFALFA) Survey. These authors report that the FUV-based SFRs were slightly higher than the SFRs from SED fitting, in general agreement with the results from a larger sample of more massive galaxies studied by \citet{Salim2007}.

\section{The Nature of the Dispersion in FUV-based SFRs\label{nature_dispersion}}
\begin{figure}
\includegraphics[width=0.48\textwidth]{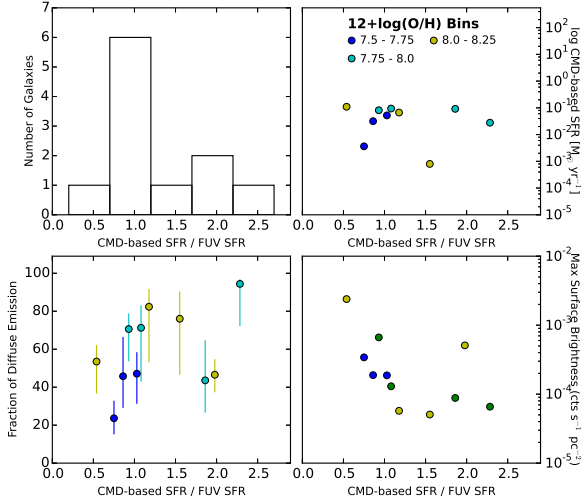}
\caption{\textit{Top left:} Histogram of the ratio of CMD-based to FUV-based SFRs using the new calibration. \textit{Top right:} The CMD-based SFRs versus the ratio of the CMD-based SFRs to FUV-based SFRs. The points are color-coded by metallicity as described in the legend. The is no correlation seen between the dispersion in the SFR indicators and the SFR. \textit{Bottom left:} The percent of diffuse FUV emission versus the ratio of CMD-based SFRs to FUV-based SFRs. Uncertainties represent the difference in diffuse emission if the empirically selected threshold delineating diffuse from UV-bright knots is varied by $\pm50$\%. There are no galaxies with a high fraction of diffuse emission with lower CMD-based to FUV-SFR ratios or vice-versa. \textit{Bottom right:} Maximum surface brightness of the galaxies versus the ratio of CMD-based SFRs to FUV-based SFRs. Similar to the previous panel, there are no galaxies with higher surface brightness values and high CMD-based to FUV-based SFRs, or vice-versa.}
\label{fig:dependency}
\end{figure}

The CMD-based SFRs and extinction corrected FUV luminosities shown in Figure~\ref{fig:calibration} are clearly correlated but there is still a high degree of dispersion. The uncertainties in the CMD-based SFRs and the statistical uncertainties in the FUV luminosities, which are approximately the size of the points, generally do not overlap with the best-fit line used for the calibration. We plot the differences between the CMD-based SFRs with the FUV-based SFRs calculated using our new calibration in the upper left panel of Figure~\ref{fig:dependency}. For clarity, we plot only the galaxies used in the SFR calibration (see Section~6). The ratios of the two SFR indicators reveal that there is still up to a factor of two difference between the measurements. 

As discussed in Section~6, most of this dispersion appears to be intrinsic and can be due to a number of different factors. The more likely factors include (i) stochastic effects on the upper-end of the IMF which depends on the level of star-formation activity, (ii) the known dependency of the FUV flux on stellar metallicity, and (iii) the amount of concentrated versus diffuse FUV emission. In Figure~\ref{fig:dependency}, we investigate the dependency of the dispersion in the SFRs on these three factors using measurable quantities, namely the level of star-formation activity, gas-phase metallicity measurements, and the amount of concentrated versus diffuse FUV emission. 

In the top right panel of Figure~\ref{fig:dependency}, we plot the ratio of CMD-based SFR to FUV-SFR calculated using our new calibration constant against the CMD-based SFRs. Based on the premise that higher SFRs will be less susceptible to stochastic effects, the expectation is that there will be an decrease in dispersion at higher SFRs. However, there is no clear correlation between the dispersion and the star-formation activity, suggesting that this is not the dominate driver of the dispersion. Note that this does not imply that stochastic effects do not contribute to the scatter. As demonstrated in Section~4.2, the FUV flux predictions from the {\tt MATCH} and SLUG models $-$ which stochastically populate the IMF $-$ have the lowest scatter of the models tested. 

Second, we consider the known dependency the metallicity of a star has on the FUV emission. At a given SFR, there is less UV flux produced from higher metallicity stellar populations due to lower stellar temperatures and line blanketing. Thus, when a constant scaling relation is applied over a range in metallicity, there is an increasing deviation between the input and recovered SFRs as a function of metallicity. Previous authors have reported on the amplitude of this dependency, with UV fluxes varying by less than 0.24 dex, or $\sim70$\%, over a metallicity range from $Z$ of 0.0003 to 0.03 \citep[e.g., see][]{Madau2014}. Other authors report a change of 55\% in the flux over log $Z_{\odot}/Z$ range of $+0.2$ to $-1.0$ \citep{Conroy2009}. 

Despite this known dependency, we do not find an empirical signature for the FUV-based SFR calibration to be changing as a function of metallicity. In Figure~\ref{fig:dependency}, the points are color-coded by metallicity bins with bluer colors representing galaxies that are more metal-poor and yellow colors representing galaxies that are comparatively more metal-rich (see legend for metallicity ranges). Changes in the UV flux as a function of metal content over this range are likely present in the stellar populations, yet the dispersion between the SFR indicators do not show a clear dependence on metallicity. From our Starburst99 simulations (see Section~4.2), the expected amplitude of the dependence is $\ltsimeq15$\% over the narrow metallicity range of the STARBIRDS sample. This dependency is smaller than the offset between the CMD-based SFRs and the FUV-based SFRs and, thus, is probably not the dominant driver of the dispersion.

Third, inspection of the GALEX and HST images of the galaxies suggests that the presence and number of star forming regions with concentrated UV emission (``UV-bright knots'') and the fraction of emission originating outside of such regions (``diffuse emission'') may be more directly correlated with the dispersion in the FUV-based SFRs. The UV-bright knots trace the young star-forming complexes in the galaxies; the source of diffuse UV emission is not clear. On the one hand, some of the diffuse UV light may originate from the main star forming complexes and be scattered by dust and gas within a galaxy. On the other hand, recent studies have suggested that the diffuse emission is from B stars that have migrated from stellar associations into field regions of the galaxies \citep[e.g.,][]{Cole1999, Tremonti2001, Chandar2005}. The timescale required by these studies for dispersing the B stars from the star forming associations is of order $\sim10$ Myr. However, other studies report a substantially longer timescale of $\sim200$ Myr for stellar clusters in dwarf galaxies to dissolve into the field \citep{Gieles2008, Bastian2010, Bastian2011}. This result is supported by a separate study on the cluster disruption fraction in NGC~4449 \citep{Annibali2011}. Separately, \citet{Liu2011} report that 80\% of the FUV emission originates from stars with ages $<30$ Myr compared to 20\% of the FUV emission from older, lower-mass stars that have presumably migrated from their clustered birth places. However, the actual measured fraction of emission in clusters can be as low as 60\% due to the higher levels of attenuation in the more dusty nascent regions. This can be compared with work from \citet{Johnson2013} who suggest that the mean age of the stellar population contributing to the FUV emission is strongly dependent on the SFH of a system, ranging from $\sim50$\% produced by stars younger than 16 Myr to 100\% of the flux produced by stars older than 100 Myr. 

Regardless of the origin of the diffuse emission, we investigated whether the dispersion in the ratio of SFR indicators is due to the relative contribution of UV-bright knots to diffuse emission. We quantify the overall differences in the UV emission in two different ways from the GALEX data. First, we consider how the fraction of diffuse UV emission relative to the UV-bright knots in the STARBIRDS sample correlates with the difference in FUV-based SFRs. This is an approach that is only possible for spatially resolved galaxies. Second, we compare how the maximum intensity in FUV emission correlates with the same difference, which enables an evaluation of this property for unresolved galaxies. 

In the bottom left panel of Figure~\ref{fig:dependency}, we plot the percentage of diffuse emission compared to the ratio of the CMD-based SFRs to FUV-based SFRs. A rough estimate of the fraction of diffuse emission was calculated from the ratio of the flux in all pixels less than an empirically selected threshold of 0.025 counts s$^{-1}$ pixel$^{-1}$ to the total flux in the FUV image. The threshold value was determined by examining the flux contour levels that encompass the UV-bright knots across the sample. The uncertainties represent the range in the percent of diffuse emission when the threshold value of 0.025 counts s$^{-1}$ pixel$^{-1}$ is varied by $\pm50$\%. As the UV-bright knots are sharply peaked in flux, varying this threshold does not significantly change the distribution of points.

There appears to be a loose correlation in Figure~\ref{fig:dependency} which shows that galaxies with larger amounts of diffuse emission have CMD-based SFRs that are larger than the FUV-based SFRs. On one hand, this loose correlation could be the result of changes in the SFR at slightly older times compared to the most recent 100 Myr. On the other hand, the origin of diffuse FUV light is not well understood. A detailed comparison study of the resolved stellar populations in \textit{both} the UV and the optical might help clarify the source of the emission. However, a Kolmogorov-Smirnov test rejects the null hypothesis that these two samples are drawn from the same distribution with a p-value of $<1$\%, indicating that these variables are not likely dependent upon one another.

Because it is only possible to calculate the fraction of diffuse emission in resolved galaxies, we probed whether a different tracer of UV bright knots, the maximum intensity in FUV emission, could be correlated with the dispersion in FUV-based SFRs. In the bottom right panel of Figure~\ref{fig:dependency}, we plot the maximum value of counts s$^{-1}$ pc$^{-2}$ for each of the GALEX FUV images versus the ratio of CMD-based SFRs with FUV-based SFRs. While there is no clear trend in the data, there are no galaxies with higher values of both maximum surface brightness and CMD-based to FUV-SFR ratios or vice-versa. However, similar to the previous comparison, a Kolmogorov-Smirnov test returns a a p-value of $<1$\% indicating these samples are likely drawn from different distributions.

\section{Conclusions \label{conclusions}}
We have compared the star formation traced by extinction corrected GALEX UV imaging in 19 resolved starburst and post-starburst dwarf galaxies with HST optical imaging of the resolved stellar populations using data accessible in the Panchromatic STARBIRDS Archive and described in a companion paper \citep{McQuinn2015}. The UV emission was corrected for extinction by estimating the dust content from MIPS imaging. Here we summarize our conclusions.

\begin{enumerate}
\item We find that SFHs derived from resolved stellar populations can accurately predict the integrated NUV fluxes in the GALEX bandpass using four independent models for the majority of the sample. This result provides independent confirmation of the transformation of recent SFRs to NUV fluxes from stellar evolution libraries, the robustness the SFHs based on CMD fitting methods, and the application of UV extinction corrections in our sample. In contrast, the integrated FUV fluxes predicted using the same inputs and models do not agree with the extinction correction FUV fluxes measured from the GALEX images. The offset at FUV wavelengths is likely driven by calibration differences in the stellar atmospheric models, systematic uncertainties in the stellar evolutionary isochrones, or both. 

\item For the majority of the sample, the FUV-based SFRs based on existing scaling relations are systematically \textit{lower} than the CMD-based SFRs. We show that the offset in the FUV-based SFRs cannot be explained by uncertainties of variable SFHs, mismatched emission timescales, or attenuation by dust. Additionally, while stochastic sampling of the IMF at low SFRs, particularly in highly clustered SF, introduces dispersion in the SFRs calculated using FUV-based SFR scaling relations in this SFR regime, stochastic effects cannot account for the discrepancy between the FUV-based SFRs from CMD-based SFRs. We caution that our sample is small; our results would benefit from further work on a larger sample of galaxies.

The offset between the CMD-based SFRs and integrated FUV-based SFRs is in agreement with the offset between predicted to measured FUV fluxes using three different models. The offset is likely driven by the theoretical calibrations of either the stellar atmosphere libraries or stellar evolutionary isochrones (or both) used in Starburst99 to calibrate the FUV-based SFR scaling relation and in the codes GALAXEV, SLUG, and {\tt MATCH} used to predict the FUV fluxes. 

\item Using the CMD-based SFRs from the low-metallicity galaxies that best meet the assumption of constant star formation over the past $\sim100$ Myr, we calculate a conversion from integrated GALEX FUV emission to FUV-based SFRs of $2.04\pm0.81\times10^{-28}$. This is the first empirical calibration of the FUV-based SFR scaling relation using a sample with well-measured and well-constrained star-formation properties outside of the synthesis model approach. The value of the conversion factor is 53\% larger than that of \citet{Hao2011} calibrated from the Starburst99 model. 

Because CMD-based SFRs are derived using more information from the stellar populations and thus use fewer assumptions, they are likely a truer measure of the star-formation activity than the FUV-based SFRs calculated using theoretically calibrated scaling relations. Additionally, the CMD-based SFRs use a stellar evolutionary library with transformations to the HST filter system. These transformations to optical wavelength use spectrophotometric standards and large libraries of stellar spectra \citep[][and references therein]{Girardi2008}. In contrast, the theoretical model used to calibrate the FUV-based SFR scaling relation relies on model atmospheres from a smaller number of massive stars that are not as well understood as the lower-mass stars included in the HST transformations. Further, in contrast to the theoretical calibrations, the empirical derivation allows a robust estimate of the uncertainty on the scaling factor. Thus, our empirically derived FUV-based SFR scaling relation promises an improvement over previous theoretical relations. 

\item The majority of the scatter between CMD-based SFRs and FUV emission from low-mass galaxies is intrinsic, reflecting the complex nature of these systems. While stochastic sampling of the upper-end of the IMF likely introduces some scatter into the relationship, we do not find empirical signatures of a dependency on star-formation activity driving the dispersion. Further, we do not find an empirical signature of the known dependence of FUV emission on metallicity driving the dispersion. We do note a possible dependency of the conversion factor on the fraction of diffuse FUV emission compared to UV-bright knots in a galaxy, although statistical tests showed little probability of a correlation.  A detailed understanding of the source of the diffuse emission is needed before its impact on measuring SFRs from integrated light can be further quantified. 

\end{enumerate}

\section{Acknowledgments}
Support for this work was provided by NASA through an ADAP grant (NNX10AD57G), and a NASA GALEX grant (NNX10AH98G). N.~P.~M is grateful for an NSF REU grant (PHYS~0851820) at the University of Minnesota. GALEX (Galaxy Evolution Explorer) is a NASA Small Explorer, launched in 2003 April. We gratefully acknowledge NASA's support for construction, operation, and science analysis for the GALEX mission, developed in cooperation with the Centre National d'Etudes Spatiales of France and the Korean Ministry of Science and Technology. This research made use of NASA's Astrophysical Data System, the NASA/IPAC Extragalactic Database (NED) which is operated by the Jet Propulsion Laboratory, California Institute of Technology, under contract with the National Aeronautics and Space Administration, and the HyperLeda database (http://leda.univ-lyon1.fr). Finally, we would like to thank Michel Fioc, Michele Fumagalli, Cai-Na Hao, Claus Leitherer, and Brigitte Rocca-Volmerange for helpful discussions on modeling star formation, and John Cannon and Yaron Teich for constructive feedback on the manuscript.

{\it Facilities:} \facility{GALEX Space Telescope, Hubble Space Telescope, Spitzer Space Telescope}

\appendix
\section{Possible but Unlikely Sources of the Offsets between FUV-based SFRs and CMD-based SFRs\label{unlikely_sources}}
Comparisons between SFR indicators can be difficult because each indicator is based on a variety of assumptions and can be impacted by different observational factors. For example, an FVU SFR is based on the predication that the SFR in a system has been constant over the lifetime of the FUV emitting stars. CMD-based SFRs assume an accurate accounting of stars above the completeness with minimal loss in the photometry due to blending and source confusion. Differences in attenuation in the wavelength dependent measurements introduce uncertainty in SFR comparisons and are often invoked as an explanation when optical and UV SFR indicators do not fully agree. These make up a few of the possible sources of the offset between the FUV-based SFRs and the CMD-based SFRs reported here. However, the advantage of the STARBIRDS data and this study is that many of these assumptions and differences can be explored in detail. Here, we itemize a number of possible explanations for the discrepancy between the FUV and CMD-based SFRs and discuss why these are $\textit{unlikely}$ sources of the offset seen in Figure~\ref{fig:opt_uv_sfrs}.

\begin{enumerate}
\item The deviations in FUV-based SFRs cannot be attributed to extinction correction uncertainties at the UV wavelengths. This is supported by several arguments. First, UV attenuation is expected to be low as the sample is mainly comprised of low metallicity dwarfs. Second, uncertainties due to differences in metallicity and UV extinction would more likely introduce scatter into the relationship, instead of an offset. Third, there is no correlation between the ratio of the CMD-based SFRs to FUV-based SFRs and the oxygen abundance of the galaxies. Fourth, careful accounting of the UV extinction was performed using measurements of the infrared emission from the MIPS images. Finally and most importantly, the FUV fluxes predicted from the resolved stellar populations were in good agreement with the measured, extinction corrected FUV fluxes. Since the CMD-based SFRs are not as heavily impacted by extinction as the UV fluxes, the agreement between predicted and measured extinction corrected UV fluxes independently confirms that the UV fluxes used in the SFR scaling relations are not significantly impacted by extinction. 

For the four galaxies whose FUV-based SFRs are $\textit{higher}$ than the CMD-based SFRs, we consider the possibility that we have $\textit{over}$-corrected for extinction. This could occur if the assumption that the majority of the UV flux comes from young stellar populations does not hold. If the primary source for heating the dust originates from other mechanisms, such as mechanical energy from stellar winds, galactic winds, and supernovae and/or light from somewhat older stellar populations, then using the IR emission as a measure of UV extinction will overestimate the correction. However, the extinction corrections for these low metallicity dwarfs are fairly modest. Only by removing $\textit{all}$ of the extinction corrections from the GALEX fluxes do the measured values agree with the predictions. In other words, for the FUV-based SFRs to agree with the CMD-based SFRs, $\textit{all}$ of the IR emission would have to be attributed to mechanisms other than re-processed UV light. As this would be an unphysical interpretation of the IR emission, we rule out over-correcting the GALEX fluxes for extinction as the main source of the discrepancy in these four systems.

\item The deviations in the FUV-based SFRs from the CMD-based SFRs are not dominated by under-sampling the upper-end of the IMF at low SFRs. As discussed above, while SFR scaling relations are based on the assumption of a fully-populated IMF, the upper-end of an IMF may not be fully populated at low-SFRs. In the UV continuum, this introduces an uncertainty of only $\sim$20\% for SFRs $\sim0.0003$ \msun yr$^{-1}$ \citep{Lee2011}. Further, if the lower FUV-based SFRs were a result of less high-mass stars at lower SFRs depressing the expected UV flux, there would a dependency in the offset on the SFR. The SFRs shown in Figure~\ref{fig:opt_uv_sfrs} span nearly 4 orders of magnitude, but the offset is roughly constant across the whole sample (excluding the galaxies with massive clusters previously discussed in Section~3.1.1). Thus, while stochasticity does have an effect on the measured UV fluxes as described by numerous studies \citep[e.g.,][]{Goddard2010, Lee2011, Koda2012, Johnson2013} and is likely a contributing factor to the scatter in the comparison, it does not explain the offset between SFRs measured in the sample. This is supported by our comparison between the predicted fluxes based on the CMD-based SFRs and the three separate models discussed in Section~4. In contrast to the {\tt MATCH} and SLUG models which randomly populate an IMF for a given SFR, the GALAXEV model does not. Yet all three are in agreement with one another. 

\item The offset in the FUV-based SFRs is not due to a breakdown in the assumption of constant star formation. Scaling the FUV emission to a SFR is sensitive to the assumption of constant star formation activity over the $\sim100$ Myr lifetime of the FUV emitting population of stars. When considering the uncertainties, this assumption of constant star formation is generally supported in the STARBIRDS sample. We can evaluate this assumption on a galaxy by galaxy basis. Looking at the SFHs shown in Figure~2 of \citet{McQuinn2010b}, a number of galaxies have only small variations in their SFRs over the past 100 Myr. Specifically, NGC~4068, ESO~154$-$023, NGC~784, DDO~165, NGC~6789, NGC~4163, and NGC~1569 have experienced sustained star formation activity over this timescale, yet show the largest discrepancy with the FUV-based SFRs of more than a factor of 4 below the CMD-based SFRs. A few of the galaxies, such as UGC~4483, UGC~6456, NGC~2366, and Holmberg~II, do deviate from the assumption of constant star formation over the past 100 Myr, with increases in their SFR within the most recent $\sim20-40$ Myr. The FUV-based SFRs for these galaxies are not as low relative to the CMD-based SFRs, reflecting this recent increase in SF activity. The deviation from a constant SFR actually $\textit{improves}$ the agreement between the SFRs. Note that while the CMD-based SFHs track changes in star formation over the past 100 Myr, as with any technique, the temporal sampling achievable is finite. One can imagine changes in the SFRs on shorter timescales than are recovered in the SFHs. As discussed above, such changes do occur and will introduce scatter into the comparison between the integrated FUV emission SFRs and the CMD-based SFRs. However, contrary to these fluctuations that are expected to introduce $\textit{scatter}$ to the relationship, we find an $\textit{offset}$. 

\item The offset between the FUV-based SFRs from the CMD-based SFRs are not due to differences in timescales of emission between the two wavelength regimes. There is still some debate on the timescale over which the FUV emission originates. For more massive galaxies, models suggest that 90\% of the FUV emission originates from stars less than 100 Myr old \citep[e.g.,][]{Kennicutt1998, Hao2011}. However, for low-mass galaxies there can be a significant contribution from stars older than 100 Myr to the integrated FUV flux, depending on the SFH \citep{Johnson2013}. Thus, we tested the sensitivity of the offset in FUV-based SFRs to different timescales. Specifically, we averaged the CMD-based SFR over timescales ranging from 50 - 200 Myr in increments of 50 Myr. Within the timescales probed, the offset in FUV-based SFRs was present in the same systems with only slight variations. This can be seen in Figure~\ref{fig:opt_uv_sfrs} where the uncertainties on the CMD-based SFRs are based on the standard deviation in the SFR from $50-150$ Myr. Very few of the error bars overlap with the FUV-based SFRs using the scaling relation from \citet{Hao2011}. 

\item There was no dependency in the offset in the FUV-based SFRs with inclination angle. The geometry and inclination of a galaxy may impact the comparison for individual systems. For example, FUV-based SFRs derived from scaling relations in edge-on galaxies, such as NGC~784 and ESO~145$-$023, are known to have higher uncertainties. However, the UV fluxes predicted from the CMD-based SFRs were in good agreement with the measured, extinction corrected fluxes. Thus, while it may be that \textit{both} the CMD-based and FUV-based SFRs are affected by the geometry in these galaxies, the inclination angle cannot alone explain the deviation between the SFRs.
\end{enumerate}

\clearpage
%
\begin{deluxetable}{lrrrccccr}
\tabletypesize{\tiny}
\tablewidth{0pt}
\tablecaption{Galaxy Sample and Properties\label{tab:galaxies}}
\tablecolumns{10}
\tablehead{
\colhead{}					&
\colhead{R.A.}				&
\colhead{Decl.}				&
\colhead{M$_B$}			&
\colhead{Distance}			&
\colhead{D$_{25}$}			&
\colhead{A$_R$}			&
\colhead{Oxygen}			&
\colhead{}					\\
\colhead{Galaxy}			&
\colhead{(J2000)}			&
\colhead{(J2000)}			&
\colhead{(mag)}			&
\colhead{(Mpc)}				&
\colhead{(arcmin)}			&
\colhead{(mag)}			&
\colhead{Abundance}		&
\colhead{Ref.}				\\
\colhead{(1)}				&
\colhead{(2)}				&
\colhead{(3)}				&
\colhead{(4)}				&
\colhead{(5)}				&
\colhead{(6)}				&
\colhead{(7)}				&
\colhead{(8)}				&
\colhead{(9)}				
}
\startdata
\multicolumn{9}{c}{\textbf{Starburst Galaxies}} \\
\\
UGC 4483	& 08:37:03.0s 	& $+$69:46:31s 	& $-12.73$   & 3.2  & 1.2  & 0.091    & 7.50 & 1    \\
UGC 6456	& 11:27:59.9s 	& $+$78:59:39s	& $-14.03$   & 4.3  & 1.5  & 0.096    & 7.64 & 2    \\
DDO 165	 	& 13:06:24.85s	& $+$67:42:25s	& $-15.09$   & 4.6  & 3.5  & 0.065    & 7.80 & 3     \\
IC 4662		& 17:47:08.8s 	& $-$64:38:30s		& $-15.13$   & 2.4  & 2.8  & 0.188    & 8.17 & 4    \\
NGC 6822	& 19:44:56.6s 	& $-$14:47:21s		& $-15.22$   & 0.5  & 15.5 & 0.632    & 8.11 & 5     \\
\\
NGC 4068	& 12:04:00.8s 	& $+$52:35:18s	& $-15.48$   & 4.3  & 3.2  & 0.058    & 7.84 & 3     \\
NGC 2366	& 07:28:54.6s 	& $+$69:12:57s	& $-16.02$   & 3.2  & 7.3  & 0.097    & 8.19 & 6    \\
ESO 154-023	& 02:56:50.38s	& $-$54:34:17s		& $-16.40$   & 5.8  & 8.2  & 0.045    & 8.01 & 3     \\
NGC 784		& 02:01:17.0s 	& $+$28:50:15s	& $-16.59$   & 5.2  & 6.6  & 0.158    & 8.05 & 3     \\
Ho II	 		& 08:19:04.98s& $+$70:43:12s		& $-16.72$   & 3.4  & 7.9  & 0.086    & 7.92 & 7     \\
\\
NGC 4214	& 12:15:39.2s 	& $+$36:19:37s	& $-17.19$   & 2.7  & 8.5  & 0.058    & 8.38 & 8     \\
NGC 5253        & 13:39:55.9s   & $-$31:38:24s		& $-17.38$   & 3.5  & 5.0  & 0.186    & 8.10 & 9     \\
NGC 1569	& 04:30:49.0s 	& $+$64:50:53s	& $-18.17$   & 3.4  & 3.6  & 1.871    & 8.19 & 10     \\
NGC 4449	& 12:28:11.9s 	& $+$44:05:40s	& $-18.27$   & 4.2  & 6.2  & 0.051    & 8.32 & 11     \\
\\
\multicolumn{9}{c}{\textbf{Post-Starburst Galaxies}} \\
\\
Antlia Dwarf	& 10:04:04.1s	& $-$27:19:52s		& $- 9.75$   & 1.3  & 2.0  & 0.212    & 7.39 & 12  \\
UGC 9128	& 14:15:56.5s 	& $+$23:03:19s	& $-12.71$   & 2.2  & 1.7  & 0.065    & 7.74 & 13  \\
NGC 4163	&12:12:09.1s 	& $+$36:10:09s	& $-14.24$   & 3.0  & 1.9  & 0.052    & 7.69 & 3    \\
NGC 6789	& 19:16:41.1s 	& $+$63:58:24s 	& $-14.32$   & 3.6  & 1.4  & 0.187    & 7.77 & 3    \\
NGC 625		& 01:35:04.6s 	& $-$41:26:10s		& $-16.53$   & 3.9  & 6.4  & 0.044    & 8.10 & 14  \\
\enddata

\tablecomments{Column (1) Galaxies are listed according to M$_B$ luminosity, from faintest to brightest. Columns (2) and (3) J2000 coordinates.  Column (4) M$_B$ luminosity corrected for extinction \citep{Karachentsev2004}.Column (5) Distance in Mpc. Column (6) Major axis of M$_B$ 25 mag isophote in arcmin \citep{Karachentsev2004}.  Column (7) A$_R$ (R$=$650 nm) extinction estimates are from the HI maps of \citet{Schlegel1998}.Column (8) and (9) Oxygen abundance and reference.}

\tablerefs{(1) \citet{Skillman1994}; (2) \citet{Croxall2009};  (3) L-Z relation; \citet{Lee2006}; (4) \citet{Hidalgo2001a}; (5) \citet{Hidalgo2001b}; (6) \citet{Roy1996}; (7) \citet{Lee2003}; (8) \citet{Kobulnicky1996};  (9) \citet{Kobulnicky1997a}; (10) \citet{Kobulnicky1997b}; (11) \citet{Skillman1989}; (12) \citet{Piersimoni1999} (13) \citet{vanZee1997}; (14) \citet{Skillman2003};}

\end{deluxetable}

%
\begin{deluxetable}{lrrrrr}
\tabletypesize{\scriptsize}
\tablewidth{0pt}
\tablecaption{Measured UV and IR Fluxes\label{tab:flux_lum}}
\tablecolumns{6}
\tablehead{
\colhead{}				&
\colhead{Flux$\rm{_{FUV}} \times10^{-14}$}	&
\colhead{Flux$\rm{_{NUV}} \times10^{-14}$}	&
\colhead{Flux$_{24 \micron}$}		&
\colhead{Flux$_{70 \micron}$}		&
\colhead{Flux$_{160 \micron}$}		\\
\colhead{Galaxy}			&
\colhead{(erg s$^{-1}$ cm$^{-2} \AA^{-1}$)} 	&
\colhead{(erg s$^{-1}$ cm$^{-2} \AA^{-1}$)}	&
\colhead{(Jy)}				&
\colhead{(Jy)}				&
\colhead{(Jy)}				\\
\colhead{(1)}				&
\colhead{(2)}				&
\colhead{(3)}				&
\colhead{(4)}				&
\colhead{(5)}				&
\colhead{(6)}				
}
\startdata
\multicolumn{6}{c}{\textbf{Starburst Galaxies}} \\
\\
UGC 4483      & 1.7$\pm$0.5   & 0.87$\pm$0.13 & 0.07  & 0.1   & ND   \\
UGC 6456      & 3.0$\pm$0.6   & 1.6$\pm$0.2   & 0.03  & ...   & ND   \\
DDO 165	      & 6.4$\pm$0.9   & 3.8$\pm$0.3   & ND    & 0.06  & ND   \\
IC 4662	      & ...	      & 16$\pm$1      & $>$5.0& $>9.0$& ...  \\
NGC 6822:FOV 1& 3.7$\pm$0.7   & 1.7$\pm$0.2   & 0.2   & 3.2   & 6.0  \\
NGC 6822:FOV 2& 4.4$\pm$0.8   & 2.3$\pm$0.2   & 0.01  & 3.2   & 6.9  \\
NGC 6822:FOV 3& 5.6$\pm$0.9   & 2.3$\pm$0.2   & 0.2   & 3.2   & 4.0  \\
\\
NGC 4068      & 7.5$\pm$1.0   & 4.1$\pm$0.3   & 0.02  & 0.7   & 1.0  \\
NGC 2366      & 31$\pm$2      & 17$\pm$ 1     & 0.7   & 5.0   & 3.7  \\
ESO 154-023   & 9.1$\pm$1.1   & 5.1$\pm$0.3   & 0.02  & 0.7   & 1.1  \\
NGC 784	      & 6.8$\pm$0.9   & 4.7$\pm$ 0.3  & 0.03  & 1.1   & 1.4  \\
Ho II	      & 32$\pm$2      & 17$\pm$1      & 0.2   & 2.7   & 3.0  \\
\\
NGC 4214      & 54$\pm$3      & 33$\pm$1      & 1.7   & 17    & 19   \\
NGC 5253      & 33$\pm$2      & 24$\pm$1      & 8.8   & 26    & 20.  \\
NGC 1569      & 3.2$\pm$0.7   & 2.8$\pm$0.2   & 7.4   & ...   & ...  \\
NGC 4449      & 160$\pm$5     & 85$\pm$1      & 3.1   & 41    & 71   \\
\\
\multicolumn{6}{c}{\textbf{Post-Starburst Galaxies}} \\
\\
Antlia Dwarf  & 0.33$\pm$0.22 & 0.39$\pm$0.09 & ND    & ND    & ND   \\
UGC 9128      & 1.2$\pm$0.4   & 0.82$\pm$0.13 & ND    & ND    & ND   \\
NGC 4163      & 2.5$\pm$0.5   & 1.8$\pm$0.2   & 0.003 & 0.07  & 0.2  \\
NGC 6789      & 0.8$\pm$0.3   & 0.64$\pm$0.11 & 0.008 & ...   & ...  \\
NGC 625	      & 9.7$\pm$1.2   & 6.2$\pm$0.4   & 0.8   & 5.6   & 5.8  \\

\enddata

\tablecomments{Column (1) Galaxy name. Columns (2)-(3) Integrated FUV and NUV flux measurements from the GALEX images in the HST field of view with foreground and background contamination masked. Uncertainties assume a Poisson distribution and are based on the square root of the number of counts at each wave band. Columns (4)-(6) MIPS flux measurements in the HST field of view after any evident foreground and background contamination has been masked. ND is a non-detection at that wavelength. Lower limits on the flux are noted for IC~4662 where the MIPS observations did not cover the HST field of view.}

\end{deluxetable}

\begin{turnpage}
\begin{deluxetable}{lcccccccccccc}
\tabletypesize{\tiny}
\tablecolumns{12}
\tablewidth{0pt}
\tablecaption{Measured UV Luminosities, Extinction Corrections, and Calculated FUV SFRs \label{tab:lum_sfrs}}
\setlength{\tabcolsep}{0.05in}
\tablehead{
\colhead{}			      		&
\colhead{log L$\rm{_{FUV}}$}	&
\colhead{log L$\rm{_{NUV}}$}	&
\colhead{FUV}				&
\colhead{NUV}				&
\colhead{FUV-NUV}			&
\colhead{A$\rm{_{FUV}}$}		&
\colhead{A$\rm{_{NUV}}$}		&
\colhead{log L$^0_{FUV}$}	&
\colhead{log L$^0_{NUV}$}	&
\colhead{log FUV SFR}		&
\colhead{log CMD SFR}		&
\colhead{log FUV SFR}		\\
\colhead{Galaxy}			&
\colhead{(ergs s$^{-1}$)}		&
\colhead{(ergs s$^{-1}$)}		&
\colhead{(mag)}			&
\colhead{(mag)}			&
\colhead{(mag)} 			&
\colhead{(mag)} 			&
\colhead{(mag)} 			&
\colhead{(ergs s$^{-1}$)}		&
\colhead{(ergs s$^{-1}$)}		&
\colhead{(\msun\ yr$^{-1}$)}	&
\colhead{(\msun\ yr$^{-1}$)}	&
\colhead{(\msun\ yr$^{-1}$)}	\\
\colhead{(1)}				&
\colhead{(2)}				&
\colhead{(3)}				&
\colhead{(4)}				&
\colhead{(5)}				&
\colhead{(6)}				&
\colhead{(7)}				&
\colhead{(8)}				&
\colhead{(9)}				&
\colhead{(10)}				&
\colhead{(11)}				&
\colhead{(12)}				&
\colhead{(13)}				
}
\startdata
\multicolumn{11}{c}{\textbf{Starburst Galaxies}} \\
\\
UGC~4483 	   & 40.51 & 40.38  &  16.09$\pm0.31$   & 16.02$\pm0.17$   &  0.07  & 0	 	& 0	     	&   40.51 & 40.38  &  $-2.66$ 	& $-2.43\pm0.02$ & $-2.47\pm0.21$ \\
UGC~6456 	   & 41.01 & 40.90  &  15.49$\pm0.23$   & 15.36$\pm0.12$   &  0.13  &$\geq$0	&$\geq$0	&   41.01 & 40.90  &  $-2.16$	& $-2.07\pm0.02$ & $-1.98\pm0.20$ \\
DDO~165 	   & 41.39 & 41.33  &  14.67$\pm0.16$   & 14.43$\pm0.08$   &  0.23  & 0	 	& 0.25     	&   41.39 & 41.43  &  $-1.77$	& $-1.20\pm0.02$  & $-1.59\pm0.18$ \\
IC~4662 	   	   & ..        & 41.39  &  ...     		     & 12.87$\pm0.04$   &   ...  	 &$\geq$0.50&$\geq$0.85	&   ...   & 41.73  &  ...    	& $-1.52\pm0.01$  & ...	 \\
NGC~6822 (1)     & 39.22 & 39.05  &  15.28$\pm0.21$   & 15.31$\pm0.12$   &  -0.03 & 1.54 	 & 1.38    	&   39.84 & 39.61  &  $-3.33$	& $-2.86\pm0.04$ & $-3.14\pm0.17$ \\
NGC~6822 (2)     & 39.30 & 39.18  &  15.07$\pm0.19$   & 14.98$\pm0.10$   &  0.09  & 1.41 	 & 1.17    	&   39.87 & 39.65  &  $-3.30$	& $-2.80\pm0.05$ & $-3.11\pm0.17$ \\
NGC~6822 (3)     & 39.41 & 39.20  &  14.82$\pm0.17$   & 14.96$\pm0.10$   &  -0.14 & 1.11 	 & 1.02    	&   39.85 & 39.60  &  $-3.32$	& $-3.36\pm0.05$ & $-3.13\pm0.17$ \\
\\
NGC~4068 	   & 41.40 & 41.32  &  14.50$\pm0.15$   & 14.32$\pm0.08$   &  0.18  & 0.12 	 & 0.02 	&   41.62 & 41.33  &  $-1.72$	 & $-1.47\pm0.02$ & $-1.53\pm0.18$ \\
NGC~2366 	   & 41.77 & 41.67  &  12.95$\pm0.07$   & 12.80$\pm0.04$   &  0.15  & 0.40  	& 0.24    	&   41.92 & 41.76  &  $-1.25$	& $-1.30\pm0.02$  & $-1.06\pm0.17$ \\
ESO~154$-$023 & 41.75 & 41.66  &  14.29$\pm0.13$   & 14.12$\pm0.07$   &  0.18  & 0.07 	 & 0       	&   41.77 & 41.66  &  $-1.40$	& $-1.11\pm0.01$  & $-1.21\pm0.18$ \\
NGC~784 	   & 41.53 & 41.54  &  14.61$\pm0.16$   & 14.18$\pm0.07$   &  0.43  & 0.36 	 & 0.11    	&   41.67 & 41.59  &  $-1.50$	& $-1.09\pm0.02$ & $-1.31\pm0.18$ \\
Holmberg~II 	   & 41.83 & 41.72  &  12.91$\pm0.07$   & 12.80$\pm0.03$   &  0.12  & 0.07 	 & 0       	&   41.86 & 41.72  &  $-1.31$	& $-1.12\pm0.02$ & $-1.12\pm0.17$ \\
\\
NGC~4214 	   & 41.86 & 41.82  &  12.35 $\pm0.06$  & 12.06$\pm0.03$   &  0.29  & 0.78 	 & 0.49    	&   42.17 & 42.02  &  $-1.00$	& $-1.12\pm0.02$  & $-0.81\pm0.17$ \\
NGC~5253 	   & 41.87 & 42.91  &  12.87$\pm0.07$   & 12.40$\pm$0.03   &  0.48  & 1.63 	 & 1.14    	&   42.53 & 42.36  &  $-0.64$	& $-1.13\pm0.02$ & $-0.46\pm0.17$ \\
NGC~1569 	   & 40.83 & 40.94  &15.43 $\pm0.23$    & 14.76$\pm0.09$   &0.67    &$\geq$2.91&$\geq$2.19& 41.99	&41.82  &  $-1.18$ & $-0.66\pm0.01$ & $-0.99\pm0.17$ \\
NGC~4449 	   & 42.70 & 42.61  &  11.20 $\pm0.03$  & 11.05$\pm0.02$   &  0.15  & 0.74 	 & 0.53   	&   43.00 & 42.82  &  $-0.17$	& $-0.41\pm0.01$  & $0.02\pm0.17$ \\ 
\\
\multicolumn{11}{c}{\textbf{Post-Starburst Galaxies}} \\
\\
Antlia 		   & 39.02 & 39.27  &  17.89$\pm0.71$   & 16.88$\pm0.25$   &  1.01  & ...  	& ...  		&   39.02 & 39.27  &  $-4.15$	 & $-3.96\pm0.15$  & $-3.97\pm0.33$ \\
UGC~9128 	   & 40.03 & 40.03  &  16.48$\pm$0.37   & 16.08$\pm0.17$   &  0.39  & ...  	& ...	      	&   40.03 & 40.03  &  $-3.14$	& $-2.76\pm0.06$  & $-2.95\pm0.23$ \\
NGC~4163 	   & 40.61 & 40.63  &  15.69$\pm0.26$   & 15.24$\pm0.12$   &  0.44  & 0		& 0	       	&   40.61 & 40.64  &  $-2.55$	& $-2.40\pm0.04$  & $-2.37\pm0.20$ \\
NGC~6789 	   & 40.29 & 40.35  &  16.89$\pm0.45$& 16.34$\pm0.19$      &  0.55  &$\geq$0 &$\geq$0 &   40.29 & 40.35  &  $-2.88$	& $-2.39\pm0..02$ & $-2.69\pm0.25$ \\
NGC~625 	   & 41.43 & 41.41  &  14.22$\pm0.13$   & 13.89$\pm0.06$   &  0.34  & 1.21	& 0.85    	&   41.91 & 41.75  &  $-1.26$	& $-1.56\pm0.02$  & $-1.07\pm0.17$ \\
\enddata

\tablecomments{Column 1$-$ Galaxy name. Columns 2$-$3 log of FUV and NUV luminosities calculated from the fluxes in Table~\ref{tab:flux_lum} using Equations~\ref{eq:fuv_lum} and \ref{eq:nuv_lum}. Columns 4$-$6 GALEX FUV and NUV magnitudes and FUV-NUV colors based on the measured fluxes with no corrections for extinction. Columns 7$-$8 FUV extinction based on TIR fluxes estimated from the measured Spitzer MIPS fluxes reported in \citet{McQuinn2015} using the model of the TIR/FUV flux ratio from \citet{Buat2005} and the empirical relation given by \citet{Dale2002}. See Section \ref{uv_lum} for a full description. The UV luminosities for Antlia and UGC~9128 were not adjusted as both systems were not detected in any of the MIPS images. Column 9$-$10 GALEX FUV and NUV luminosities based on the measured measured \citep{McQuinn2015} and corrected for extinction using the values in Columns 4 and 5. Column 11$-$ SFRs calculated using the relation from \citet{Hao2011} and the extinction corrected L$^0\rm{_{FUV}}$ in Column 9. Column 12$-$ Average SFRs over the last 100 Myr derived from resolved stellar populations using CMD fitting techniques. Column 12$-$ SFRs calculated using the relation from this study (Equation~\ref{eq:sfr_calibration}) and the extinction corrected L$^0\rm{_{FUV}}$ in Column 9. }

\end{deluxetable}

\end{turnpage}

%
\begin{deluxetable}{lcccccccccc}
\tabletypesize{\scriptsize}
\tablewidth{0pt}
\setlength{\tabcolsep}{0.05in}
\tablecaption{Comparison of Measured UV Luminosities with Predictions \label{tab:galex_models}}
\tablecolumns{9}
\tablehead{
\colhead{}				&
\colhead{}				&
\multicolumn{4}{c}{\textbf{FUV}}	&
\colhead{}				&
\multicolumn{4}{c}{\textbf{NUV}}	\\
\colhead{}				&
\colhead{GALEX}			&
\colhead{$\textsc{P\'{E}GASE}$/}	&
\colhead{MATCH/}				&
\colhead{SLUG/}				&
\colhead{GALAXEV/}			&
\colhead{GALEX}				&
\colhead{$\textsc{P\'{E}GASE}$/}	&
\colhead{MATCH/}				&
\colhead{SLUG/}				&
\colhead{GALAXEV/}			\\
\colhead{Galaxy}				&
\colhead{L$^0_{\rm{FUV}}$}		&
\colhead{GALEX}			&
\colhead{GALEX}			&
\colhead{GALEX}			&
\colhead{GALEX}			&
\colhead{L$^0_{\rm{NUV}}$}		&
\colhead{GALEX}			&
\colhead{GALEX}			&
\colhead{GALEX}			&
\colhead{GALEX}			\\
\colhead{(1)}				&
\colhead{(2)}				&
\colhead{(3)}				&
\colhead{(4)}				&
\colhead{(5)}				&
\colhead{(6)}				&
\colhead{(7)}				&
\colhead{(8)}				&
\colhead{(9)}				&
\colhead{(10)}				&
\colhead{(11)}				
}
\startdata
\multicolumn{11}{c}{\textbf{Starburst Galaxies}} \\
\\
UGC~4483	&	3.25E+40	&	 0.69 	&	 1.91 	&	 1.31 	&	 1.17 	&	2.41E+40	&	 1.28 	&	 1.38 	&	 1.19 	&	 0.85 	\\
UGC~6456	&	1.02E+41	&	 0.37 	&	 1.38 	&	 1.23 	&	 0.82 	&	8.01E+40	&	 0.66 	&	 0.97 	&	 1.05 	&	 0.56 	\\
DDO~165	&	2.48E+41	&	 0.60 	&	 1.99 	&	 2.22 	&	 3.41 	&	2.68E+41	&	 0.82 	&	 0.98 	&	 1.38 	&	 1.68 	\\
IC~4462	&	...	&	 ... 	&	 ... 	&	 ...	&	 ...	&	5.43E+41	&	 0.58 	&	 0.41 	&	 0.70 	&	 0.49 	\\
NGC~6822	&	7.13E+39	&	 1.97 	&	 4.75 	&	 2.69 	&	 3.41 	&	4.19E+39	&	 5.04 	&	 6.06 	&	 3.09 	&	 3.47 	\\
\\																					
NGC~4068	&	2.83E+41	&	 0.43 	&	 1.18 	&	 1.45 	&	 1.83 	&	2.12E+41	&	 0.81 	&	 0.92 	&	 1.30 	&	 1.27 	\\
NGC~2366	&	8.40E+41	&	 0.64 	&	 0.97 	&	 0.94 	&	 0.69 	&	5.81E+41	&	 1.23 	&	 0.78 	&	 0.92 	&	 0.60 	\\
ESO~154$-$023	&	5.94E+41	&	 0.36 	&	 1.01 	&	 1.44 	&	 1.93 	&	4.62E+41	&	 0.69 	&	 0.70 	&	 1.24 	&	 1.38 	\\
NGC~784	&	4.66E+41	&	 0.45 	&	 1.29 	&	 1.93 	&	 2.64 	&	3.86E+41	&	 0.84 	&	 0.88 	&	 1.57 	&	 1.77 	\\
Holmberg~II	&	7.27E+41	&	 0.46 	&	 1.45 	&	 1.26 	&	 1.08 	&	5.31E+41	&	 0.88 	&	 1.09 	&	 1.16 	&	 0.85 	\\
\\																					
NGC~42114	&	1.48E+42	&	 0.31 	&	 0.39 	&	 1.05 	&	 0.96 	&	1.04E+42	&	 0.61 	&	 0.36 	&	 1.00 	&	 0.80 	\\
NGC~5253	&	3.35E+42	&	 0.32 	&	 0.22 	&	 0.49 	&	 0.37 	&	2.31E+42	&	 0.63 	&	 0.20 	&	 0.48 	&	 0.30 	\\
NGC~1569	&	9.80E+41	&	 0.47 	&	 1.26 	&	 1.54 	&	 3.80 	&	6.57E+41	&	 1.03 	&	 1.20 	&	 1.55 	&	 3.30 	\\
NGC~4449	&	1.00E+43	&	 0.58 	&	 0.37 	&	 0.75 	&	 0.64 	&	6.60E+42	&	 1.16 	&	 0.34 	&	 0.77 	&	 0.57 	\\
\\																					
\multicolumn{11}{c}{\textbf{Post-Starburst Galaxies}} \\																					
\\																					
Antlia	&	1.05E+39	&	 1.21 	&	 1.80 	&	 1.65 	&	 0.70 	&	1.86E+39	&	 0.97 	&	 0.67 	&	 0.63 	&	 0.23 	\\
UGC~9128	&	1.07E+40	&	 0.46 	&	 1.82 	&	 1.87 	&	 1.58 	&	1.08E+40	&	 0.79 	&	 1.17 	&	 1.26 	&	 0.91 	\\
NGC~4163	&	4.12E+40	&	 0.26 	&	 1.14 	&	 1.12 	&	 1.52 	&	4.33E+40	&	 0.42 	&	 0.65 	&	 0.72 	&	 0.75 	\\
NGC~6789	&	1.96E+40	&	 0.49 	&	 1.71 	&	 2.10 	&	 4.65 	&	2.26E+40	&	 0.74 	&	 0.85 	&	 1.23 	&	 2.03 	\\
NGC~625	&	8.19E+41	&	 0.07 	&	 0.16 	&	 0.34 	&	 0.72 	&	5.57E+41	&	 0.17 	&	 0.15 	&	 0.34 	&	 0.58 	\\
\enddata

\tablecomments{Column 1$-$Galaxy name. Column 2$-$GALEX FUV band luminosity (ergs s$^{-1}$) based on the FUV fluxes measured in the equivalent HST field of view and corrected for extinction. Column 3$-$6 Ratio of the predicted FUV luminosity to the measured FUV luminosity based on SFRs(t,Z) derived from optically resolved stellar population and the synthetic spectral model $\textsc{P\'{E}GASE 2.0}$ \citep{Fioc1997}, synthetic stellar populations generated from {\tt MATCH} \citep{Dolphin2002}, the stochastic model SLUG \citep{daSilva2012}, and the synthetic galaxy spectrum from the GALAXEV code \citep{Bruzual2003}. Columns $7-11$ are the same as Columns $2-6$ but for the GALEX NUV band pass. Note the luminosities and predictions listed for NGC~6822 are the average luminosities over three fields of view.}
\end{deluxetable}

%
\begin{deluxetable}{ll}
\tablewidth{0pt}
\tablecaption{Stellar Atmosphere Models \label{tab:stellar_atm_models}}
\tabletypesize{\scriptsize}
\tablecolumns{2}
\tablehead{
\colhead{Model}				&
\colhead{Stellar Library References}	
}
\startdata
Starburst99		& \citet{Lejeune1997, Lejeune1998} for plan-parallel atmospheres which includes models from \citet{Kurucz1992}\\
				& \citet{Hillier1998} atmospheres for stars with strong winds \\	
				& \citet{Pauldrach2001} models for O star atmospheres\\
$\textsc{P\'{e}gase}$	& \citet{Kurucz1992} stellar spectra compiled by \citet{Lejeune1997, Lejeune1998} for plane-parallel atmospheres\\
{\tt MATCH}		&  \citet{Castelli2004} \\
				& Black-body spectrum for non-white dwarf stars with T$>50 000$ K \\
SLUG	  		&   \citet{Lejeune1997, Lejeune1998} for plane-parallel atmospheres based on the original model spectra from \citet{Kurucz1992}\\
				& \citet{Smith2002} implementation of atmospheres for O stars from \citet{Hillier1998}\\
			& \citet{Maeder1990} for stellar winds \\	
GALAXEV			&   \citet{Lejeune1997, Lejeune1998} for plane-parallel atmospheres which includes models from \citet{Kurucz1992}\\
				& \citet{Westera2001,Westera2002} for semi-empirical calibrations of model-atmospheres at non-solar metallicities\\

\enddata

\end{deluxetable}

\end{document}